%% file: paper.tex
\documentclass[a4paper]{scrbook} 
\usepackage{etex}
\usepackage{a4wide}
\usepackage{amsmath,amssymb,amsfonts,amsthm,makeidx,graphicx,xcolor}
\usepackage{txfonts}
\usepackage{helvet}
\usepackage{bm}
\usepackage[numbers]{natbib}
\usepackage[pdftex,pdftitle={Modeling Approaches and Computational
  Methods for Particle-laden Turbulent Flows (book chapter)},
breaklinks=true,bookmarks=true,colorlinks=true,linkcolor=blue,
citecolor=blue,allcolors=blue]{hyperref}
\usepackage{minitoc}
\usepackage[symbol]{footmisc}
\newif\ifnewauthor
\newauthortrue
\newcounter{affils}
\makeatletter
\def\chapterauthor[#1]#2{\ifnewauthor\else, \fi
  {\bfseries #2}\textsuperscript{\hyperref[authaffil\the\value{chapter}.#1]{#1}}%
  \ifnewauthor\newauthorfalse\gdef\chapterauthors{#2}\else
    \g@addto@macro\chapterauthors{, #2}\fi
  \ignorespaces 
}
\makeatother
\def\chapteraffil[#1]{%
\item[$^{#1}$]%
  \refstepcounter{affils}%
  \label{authaffil\the\value{chapter}.#1}%
}
\newenvironment{affils}
{\addtocontents{toc}{\chapterauthors\string\par}\begin{enumerate}}
  {\end{enumerate}\global\newauthortrue}
\newenvironment{abstract}{\rightskip1in 
{\bfseries Abstract:} }{}
\usepackage{scrlayer-scrpage}
\automark{chapter}
\lehead{
  Modelling approaches and computational methods for particle-laden turbulent flows}
\rohead{\leftmark}
\newenvironment{acknowledgements}%
{\rightskip1in 
{\bfseries Acknowledgements:} }{}
\usepackage{ulem}

\def\solid{\protect\rule[1pt]{10.pt}{1pt}}

\definecolor{darkgreen}{RGB}{0,102,0}
\newcommand{\revision}[2]{#2}
\newcommand{\revisions}[1]{}
\newcommand{\DPP}[2]{\frac{\partial{#1}}{\partial{#2}}}
\newcommand{\DCP}[2]{\frac{d{#1}}{d{#2}}}
\newcommand{\grad}[1]{\nabla{#1}}
\newcommand{\Div}[1]{\nabla\cdot{#1}}
\newcommand{\refeq}[1]{\mbox{{\scshape Eq}~(\ref{#1})}}

\def\Om{\Omega}
\def\bl{\bm{\lambda}}
\def\P{\mathcal{P}}
\def\bom{\bm{\omega}}
\def\bxi{\bm{\xi}}
\def\bal{\bm{\alpha}}
\def\bu{\bm{u}}
\def\bv{\bm{v}}
\def\bzero{\bm{0}}
\def\bx{\bm{x}}
\def\bU{\bm{U}}
\def\bV{\bm{V}}
\def\bg{\bm{g}}
\def\bJ{\bm{J}}
\def\bF{\bm{F}}
\def\bef{\bm{f}}
\def\bR{\bm{R}}
\def\br{\bm{r}}
\def\bD{\bm{D}}
\def\bT{\bm{T}}
\begin{document}
%
\include{chap1_Numerics}

%
\end{document}

%% file: chap1_Numerics.tex
  \setcounter{chapter}{4} 
  \chapter[Particle-resolved DNS methods]
  {Efficient methods for particle-resolved
    direct numerical simulation\footnote{%
      Published as Chapter~5 in {\itshape
        Modeling Approaches and Computational Methods for
        Particle-laden Turbulent Flows 
      },
      S.\ Subramaniam and S.\ Balachandar (editors), Academic press, 
      2022, 
      \url{https://doi.org/10.1016/B978-0-32-390133-8.00013-X}.
    }
  }
  \label{PRM_chap5} 
  %
  %
  \chapterauthor[1]{Markus Uhlmann} 
  \chapterauthor[2]{Jos Derksen}%
  \chapterauthor[3,4]{Anthony Wachs}%
  \chapterauthor[5]{Lian-Ping Wang}%
  \chapterauthor[1]{Manuel Moriche}%
  %
  \begin{affils}
  \chapteraffil[1]{%
    {Karlsruhe Institute of Technology},
    {Institute for Hydromechanics},
    {Kaiserstr.\ 12, 76131 Karlsruhe, Germany}
  }
  \chapteraffil[2]{%
    {University of Aberdeen},
    {School of Engineering},
    {King's College, Aberdeen AB24 3UE, United Kingdom}
  }
  \chapteraffil[3]{%
    {University of British Columbia},
    {Department of Mathematics},
    {1984 Mathematics Road, Vancouver BC V6T 1Z2, Canada}
  }
  \chapteraffil[4]{%
    {University of British Columbia},
    {Department of Chemical and Biological Engineering},
    {2360 East Mall, Vancouver BC V6T 1Z3, Canada}
  }
  \chapteraffil[5]{%
    {Southern University of Science and Technology},
    {Department of Mechanics and Aerospace Engineering},
    {Shenzen, Guangdong, China}
  }
  %
  %
  %
  \end{affils}
  %
  %
  %
  \minitoc
  \input{chap1_abstract}
  %
  %
  %
\input{chap1_intro}
%
\input{chap1_coupling_ibm}
\input{chap1_coupling_dlm}

\input{chap1_coupling_mesoscopic}

%
%
\input{chap1_reference_data}
\input{chap1_comparison}
%
%
%
\input{chap1_conclusions}

\begin{acknowledgements}
  {This work has received financial support from the Deutsche
    Forschungsgemeinschaft (DFG, German Research Foundation) through
    project UH~242/11-1.
  }
\end{acknowledgements}
\bibliographystyle{apa-good-doi}
%
\bibliography{TPFbook,MU1.bib,AW1.bib,LPW1.bib,JD1.bib}
%

%% file: chap1_abstract.tex
\begin{abstract}
  In the present chapter we focus on the fundamentals of
  non-grid-conforming numerical approaches to simulating particulate
  flows, implementation issues and grid convergence vs.\ available 
  reference data.
  The main idea is to avoid adapting the mesh (and -- as much as
  possible -- the discrete operators) to the time-dependent
  fluid domain with the aim to maximize computational efficiency. 
  We restrict our attention to
  spherical particle shapes (while deviations from sphericity
  are treated in a subsequent chapter).
  We show that similar ideas can be successfully implemented in a
  variety of underlying fluid flow solvers, leading to powerful
  tools for the direct numerical simulation of large particulate
  systems.  
\end{abstract}

%% file: chap1_intro.tex
\section{Introduction} %
\label{sec:PRM-intro}
Particle-resolved direct numerical simulation (PR-DNS) involves
the accurate solution of the Navier-Stokes equations for all relevant
flow scales in the presence of immersed, moving particles.
This includes both: the entire range of the a priori existing
turbulent flow features, as well as the details of the disturbance
flows induced by the particles. 
Despite its formidable computational complexity,  
the PR-DNS approach is now firmly established as a high-fidelity data
source which can serve as an alternative or as a complement to
laboratory experiments. Since the first pioneering studies 
of this type which have emerged in the late 1990s, the spectrum of
flow classes and added physics which can be tackled with PR-DNS has
continuously broadened, and the range of scales that can be faithfully
described has grown steadily. This development is owed to the continuing
increase in available computing capacity coupled with 
advances in the efficiency of the underlying numerical methods. While
very early studies were limited to a small number of particles
(e.g.\ \citet{pan:97} tracked up to 160 moving particles in turbulent
channel flow), it is nowadays feasible to describe ${\cal O}(10^6)$
\revision{particles with resources available at many HPC centers. }{%
  particles with resources available at many HPC centers
  \citep{kidanemariam:16a}.} 
An overview of the physical results which have been extracted from
PR-DNS studies to date is given in chapter~6 of
the present volume. 

In the context of mesh-based numerical methods, there exist two general
paradigms for simulating the flow around moving solid objects:
boundary-conforming and non-confoming methods. In the former case, the
grid is adapted to the time-dependent position of the
fluid-solid interface \citep{hu:01,johnson:99a}. This requires 
frequent remeshing steps as well as the corresponding mapping of
field quantities to the updated grid, both of which entail a
relatively high computational cost. A detailed account of the
boundary-conforming approach can be found in the recent review by
\citet{wachs:19}. 
In non-boundary-conforming methods, on the other hand, the fluid-solid
interface condition is imposed by other means, such as specifically
tailored source terms, modification of the discrete operators, or
introduction of Lagrange multipliers. 
At the same time the computational mesh does not need to be adapted
over time, which allows for simple data structures and extremely efficient
algorithms from single-phase flow to be applied to the particulate
flow problem with relatively few modifications. As a consequence,
the fixed-grid approach has become the method of choice for PR-DNS. 

In the present chapter we are reviewing several fixed-grid methods
which have proved highly productive in the context of turbulent
particulate flow, namely the immersed boundary method
(\S~\ref{sec:PRM-numerics-ibm}),
\revision{Lattice-Boltzmann-based methods
  (\S~\ref{sec:PRM-numerics-lbm}),
  discrete Lagrange multiplier methods
  (\S~\ref{sec:PRM-numerics-dlm}), 
  and the discrete unified gas kinetic scheme
  (\S~\ref{sec:PRM-numerics-dugks}).}{%
  distributed Lagrange multiplier methods
  (\S~\ref{sec:PRM-numerics-dlm}), 
  and Boltz\-mann-equation based mesoscopic methods
  (\S~\ref{sec:PRM-numerics-lbm}). 
}
\revision{Our focus is to give an overview of those techniques which
  are deemed most relevant to PR-DNS.}{%
  Our focus is to give an overview of those aspects of the numerical
  approaches which are deemed most relevant to PR-DNS.
} 
\revision{Due to space restrictions, however, not all methods which
  are currently in use can be covered here.}{%
  Due to space restrictions, however, not all methods which today are  
  being successfully used in the community for these types of
  applications can be covered here. 
}
Some notable omissions in this text are listed in the following: 
sharp interface and cut-cell methods \citep{udaykumar:01,schneiders:16},
volume penalization \citep{angot:99,kolomensky:09},
overset-grids \citep{burton:02,henshaw:08},
and coupling to a local Stokes flow solution \citep{zhang:03}. 
\revision{}{%
  Furthermore, a comprehensive comparison of these different methods
  is not undertaken here, since it would require a significant
  collaborative effort (cf.\ discussion in \S~\ref{sec:PRM-numerics-comp}). 
}
It should also be noted that excellent previous reviews have been
published on immersed boundary methods in a general context 
\citep{iaccarino:05},
on immersed boundary methods designed spcifically for particulate flow
\citep{haeri:12},
on the immersed boundary method for fluid-structure interaction
\citep{sotiropoulos:14,kim:19},
and on particle-resolved methods for complex-shaped bodies
\citep{wachs:19}. 
%
%
%
%
%
%
%
%
%
%
%
%
%
%
%
%
%
%
%
%
%
%
%
%
%
%
%
%
%

%% file: chap1_coupling_ibm.tex
%
\section{%
  The immersed boundary method in Navier-Stokes-based
  solvers}
\label{sec:PRM-numerics-ibm}
The immersed boundary method (often abbreviated as IBM) 
goes back to the
work of 
\citet{peskin:72} (cf.\ also \cite{peskin:72b,peskin:02})
who designed it 
as a means to perform simulations of the flow around heart valves. 
The valves in the original work are treated as flexible objects which
deform and translate under the fluid load, while at the same time
exerting a force upon the fluid. 
In order to achieve this, Lagrangian marker points are attached to the
immersed object which are advected with the local fluid velocity; the
deformation of the marker point configuration then leads to internal
stresses (depending on the desired material properties) which are
spread back to the Eulerian grid. This in turn yields an additional
force term, henceforth denoted as $\mathbf{f}$, which appears on the
right-hand-side of the fluid momentum equation. 
Therefore, the following three steps are the key ingredients of an
immersed boundary method:
(i) interpolation of the velocity field from the
Eulerian grid to the Lagrangian marker positions;
(ii) computation of the stresses acting at the Lagrangian points;
(iii) spreading of the stresses back to the Eulerian grid.
Steps (i) and (iii) are closely related operations which in a
continuous formulation involve integrals over Dirac delta functions. 
For the discrete counterpart \citet{peskin:72b} devised regularized
delta functions which ensure smooth interpolation and spreading for
arbitrarily located force points (cf.\ also \cite{peskin:02}). 

Now, when the immersed solid object of interest is not flexible, but a
perfectly rigid body, the appropriate material property can be
mimicked in the original immersed-boundary context by a spring-like
behavior with large stiffness linking neighboring marker points
\cite{goldstein:93,saiki:96}. This feed-back forcing approach,
however, has the drawback of presenting severe time step restrictions
and of introducing numerical parameters which require some amount of
tuning in practice. As a consequence, feed-back forcing is typically
avoided in PR-DNS applications.  

\subsection{%
  Direct forcing immersed boundary method}
\label{sec:PRM-numerics-ibm-direct-forcing}
\citet{mohdyusof:97} (cf.\ also \cite{fadlun:00}) realized that the
feed-back procedure can be circumvented by constructing the forcing
term such that the ``desired'' velocity is directly obtained after
each time step. The basic idea can be demonstrated by writing the
Navier-Stokes momentum equation~(1.3) 
in the following simple semi-discrete form \cite{fadlun:00} 
\begin{equation}\label{eq:PRM_ibm_momentum_1}
  \frac{\mathbf{u}_f^{n+1}-\mathbf{u}_f^n}{\Delta t} =
  \mathbf{RHS}^{n+1/2}+\mathbf{f}^{n+1/2}
  \,,
\end{equation}
where the superscript indicates a discrete time level, $\mathbf{RHS}$
regroups the advective, viscous and pressure terms (which are
evaluated at some intermediate time with index $n+1/2$), 
and $\Delta t$ is the discrete time step.
Supposing that a desired velocity value $\mathbf{u}_f^{(d)}$ is
known at a grid point coinciding with the fluid-solid interface, an
explicit expression for the forcing term 
$\mathbf{f}^{n+1/2}$ can be obtained from
\eqref{eq:PRM_ibm_momentum_1} by requiring that
$\mathbf{u}_f^{n+1}=\mathbf{u}_f^{(d)}$, viz.\ \cite{fadlun:00}
\begin{equation}\label{eq:PRM_ibm_fibm_1}
  \mathbf{f}^{n+1/2}
  =
  \frac{\mathbf{u}_f^{(d)}-\mathbf{u}_f^n}{\Delta t} 
  -
  \mathbf{RHS}^{n+1/2}
  \,.
\end{equation}
Since, however, the particle surface does in general not coincide with the
Eulerian grid nodes, interpolation of some kind is necessary in order
to define the desired velocity. Various interpolation techniques have
been proposed in the past, either based upon (linear) grid-based
operations \citep{fadlun:00,kim:01}, 
or on a combination of grid-based operations and use of a regularized
delta function \cite{limaesilva:03}. In \cite{uhlmann:04} a slightly
different approach is taken: a preliminary velocity field is first
interpolated to Lagrangian marker locations on the fluid-solid
interface (as in Peskin's original method), then the force term is
computed in the spirit of \eqref{eq:PRM_ibm_fibm_1}, and finally
the force is transferred back to the Eulerian grid, again using Peskin's
original spreading operator. Since the method presented in
\cite{uhlmann:04} serves as a foundation for many subsequent
refinements, we will specify it here in more detail.

\begin{figure}
  \begin{minipage}{.3\linewidth}
    \centerline{$(a)$}
  \end{minipage}\hfill
  \begin{minipage}{.3\linewidth}
    \centerline{$(b)$}
  \end{minipage}\hfill
  \begin{minipage}{.3\linewidth}
    \centerline{$(c)$}
  \end{minipage}
  \\
  \begin{minipage}{3ex}
    \rotatebox{90}{$\delta_h^{(1)}\Delta x$}
  \end{minipage}
  \begin{minipage}{.3\linewidth}
    \includegraphics[width=\linewidth]
    {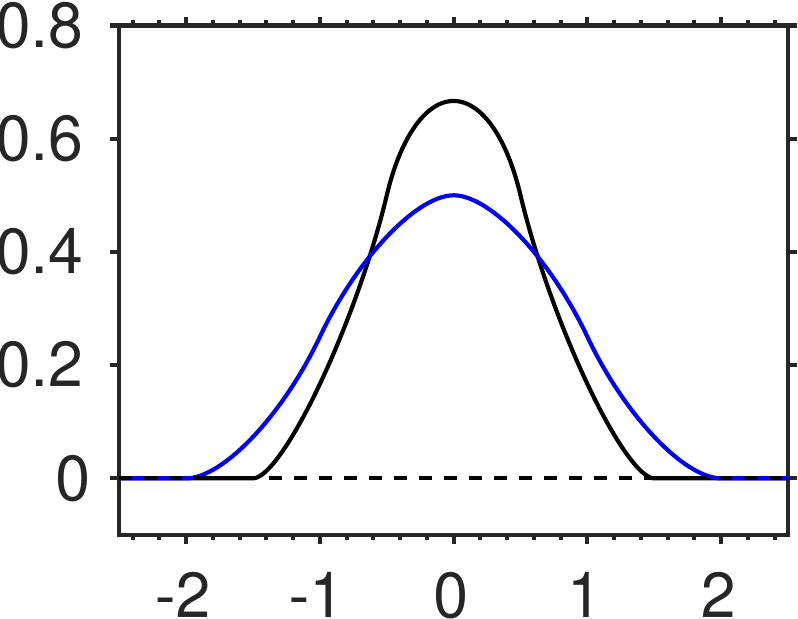}
    \\
    \centerline{$r/\Delta x$}
  \end{minipage}
  \hfill
  \begin{minipage}{.3\linewidth}
    \includegraphics[width=\linewidth]
    {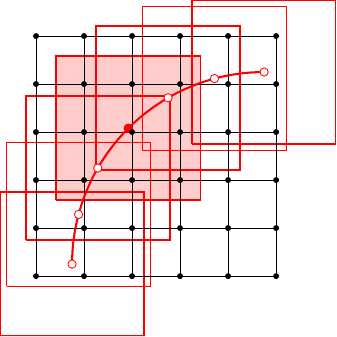}
  \end{minipage}
  \hfill
    \begin{minipage}{.2\linewidth}
    \includegraphics[width=\linewidth]
    {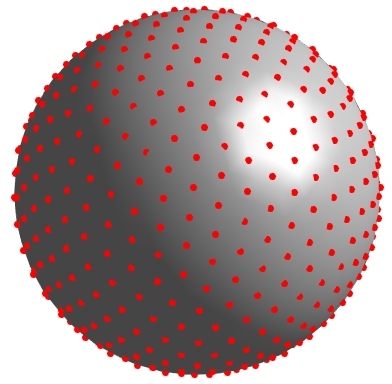}
  \end{minipage}
  \caption{%
    $(a)$
    Discrete delta function kernel $\delta_h^{(1)}(r)$ used in the
    interpolation \eqref{eq:PRM_ibm_interp_1} and spreading
    \eqref{eq:PRM_ibm_spread_1} of variables between 
    Eulerian and Lagrangian locations:
    {\color{black}\solid}~3-point function of \citep{roma:99};
    {\color{blue}\solid}~4-point function of \citep{peskin:02}.
    The three-dimensional kernel is defined as the tensor product of
    the one-dimensional kernels, viz.\ 
    $\delta_h(\mathbf{r})=\delta_h^{(1)}(r_1)\delta_h^{(1)}(r_2)\delta_h^{(1)}(r_3)$. 
    $(b)$ Sketch of a uniform and isotropic Eulerian grid (grid points
    in black) and of the Lagrangian marker points (red dots) distributed
    along a segment of a circle.
    In this case a distribution with an arc-length $\Delta s=\Delta x$
    has been chosen. 
    The red squares indicate the support of the discrete delta
    function kernel (in this case the 3-point function of
    \citep{roma:99}) associated to each Lagrangian marker point.
    $(c)$ Distribution of 515 Lagrangian points on the surface of a
    sphere (as e.g.\ used in an IBM when $d_p/\Delta x=13$), obtained
    via a repulsion force algorithm \citep{saff:97}.  
  }\label{fig:PRM_ibm_sketch_1}
\end{figure}

Let us define a set of $N_L$ Lagrangian marker points distributed over
the immersed surface of a solid object, with $\mathbf{X}_k(t)$
denoting the $k$th marker's position at time $t$ (cf.\
figure~\ref{fig:PRM_ibm_sketch_1}). Each Lagrangian marker is
associated with a forcing volume $\Delta V_k$ which is of the order of a grid
cell volume. 
The first step is to determine a preliminary velocity field
$\tilde{\mathbf{u}}_f$, solution of the momentum equation 
\eqref{eq:PRM_ibm_momentum_1} in the absence of any
immersed-boundary forcing, viz. 
\begin{equation}\label{eq:PRM_ibm_momentum_2}
  \frac{\tilde{\mathbf{u}}_f-\mathbf{u}_f^n}{\Delta t} =
  \mathbf{RHS}^{n+1/2}
  \,.
\end{equation}
Next, the field $\tilde{\mathbf{u}}_f$ is interpolated to the set of
Lagrangian marker points with the aid of the
regularized delta function $\delta_h$:
\begin{equation}\label{eq:PRM_ibm_interp_1}
  \tilde{\mathbf{U}}_f(\mathbf{X}_l)
  =
  \sum_{i,j,k}%
  \tilde{\mathbf{u}}_f(\mathbf{x}_{ijk})\,
  \delta_h(\mathbf{X}_l-\mathbf{x}_{ijk})\,\Delta x^3
  \,,\qquad\forall l\,.
\end{equation}
In \eqref{eq:PRM_ibm_interp_1} we assume that the Eulerian
grid is uniform and isotropic, i.e.\ that
\revision{$\mathbf{x}_{ijk}=\mathbf{x}_0+(i\Delta x,j\Delta x,k\Delta
  x)$.}{%
  $\mathbf{x}_{ijk}=(i\Delta x,j\Delta x,k\Delta
  x)$.
}
The specific force $\mathbf{F}$ which drives the preliminary velocity
towards the desired value $\mathbf{U}^{(d)}$ at a given Lagrangian
marker point (which can be expressed in terms of the particle's
translational and rotational velocity) is then obtained through the
following relation:  
\begin{equation}\label{eq:PRM_ibm_fibm_lag_1}
  \mathbf{F}(\mathbf{X}_l)=
  \frac{\mathbf{U}^{(d)}(\mathbf{X}_l)
      -\tilde{\mathbf{U}}_f(\mathbf{X}_l)}{\Delta t}
  \,,\qquad\forall l\,.
\end{equation}
Finally, the force at the Lagrangian positions is spread back to the
Eulerian grid,
\begin{equation}\label{eq:PRM_ibm_spread_1}
  \mathbf{f}^{n+1/2}(\mathbf{x}_{ijk})=
  \sum_{l}\mathbf{F}(\mathbf{X}_l)\,
  \delta_h(\mathbf{X}_l-\mathbf{x}_{ijk})\,\Delta V_l
  \,,\qquad\forall i,j,k
  \,,
\end{equation}
which then lets us solve the momentum
equation~\eqref{eq:PRM_ibm_momentum_1} over the entire domain 
occupied by both fluid and \revision{solids.}{solid phases.} 

The immersed boundary method as outlined in equations
\eqref{eq:PRM_ibm_momentum_1} and
(\ref{eq:PRM_ibm_momentum_2}-\ref{eq:PRM_ibm_spread_1}) is
particularly well adapted to PR-DNS as it allows for smooth variations
of hydrodynamic force and torque while particles are free to translate
with respect to the Eulerian grid.  
At the same time the sharp interface is essentially ``smeared'' over
the width of the
transfer kernel function, which has a
support of three to four grid widths in typical applications.
\revision{}{%
  One consequence of ``smearing'' the fluid-solid interface in diffuse
  interface methods is that the drag force on a particle tends to be
  overestimated (cf.\ \S~\ref{sec:PRM-numerics-ibm-accuracy} below for
  a demonstration with the aid of a scalar equation in one space
  dimension and further discussion of this issue). 
}
Here the regularized delta function as constructed by \citet{peskin:02}
guarantees that the net force and torque acting on a body are
preserved by the spreading step. This equivalence is, however, limited
to uniform Cartesian grid systems; interpolation and spreading
operators applicable to non-uniform and curvilinear grids have been
proposed by various authors
\citep{zhang:04,vanella:09,pinelli:10,akiki:16a}.  
In practice it is relatively straightforward to implement the
algorithm outlined above into different flow solvers with various
flavors of temporal and spatial discretizations.  
\subsection{%
  How and where to distribute Lagrangian marker points?} 
\label{sec:PRM-numerics-ibm-lagpoints}
Let us first consider the case where we wish to distribute
Lagrangian points over the surface of a sphere. On a uniform grid, a
computationally optimal distribution is such that the points are
``evenly'' spread in some sense, as it should require the smallest
total number of points (and hence the lowest operation count) for a
desired point density.  
Although no unique measure of ``evenness'' of a spatial distribution
in three spatial dimensions exists, both iterative approaches as well
as explicit constructions on the sphere are available for practical
purposes, as discussed e.g.\ in \cite{saff:97} and \cite{leopardi:06}. 

Since the early work of \citet{fadlun:00} many researchers have chosen
to apply an immersed boundary force only at the fluid-solid
interface, while allowing the flow field in the interior of
the solid domain to evolve freely. Although the impact of these
parasitic currents upon the flow in the actual fluid domain is
generally found to be small, their effect can in some cases become
noticeable. \citet{moriche:20a} report a case of non-physical path
oscillations due to surface forcing in their simulations of mildly
oblate spheroids with relatively small excess density settling under
gravity. 
\citet{uhlmann:05a} observed that applying the direct forcing
immersed-boundary procedure throughout the volume occupied by the
solid particle tends to reduce the slip error at the interface for low
Reynolds number flows.
Therefore, it is sometimes beneficial to collocate Lagrangian force
points throughout the solid volume,
in the spirit of DLM methods discussed in 
\S~\ref{sec:PRM-numerics-dlm},
although this entails an increase 
in the particle-related operation count. 
For this purpose additional interior layers of force points can simply
be added (akin to an onion) until the particle volume is covered.
This construction is not applicable to shapes which deviate 
strongly from a sphere. In the general case, an iterative (but still
relatively efficient) algorithm based on Vorono\"i tesselation is
available \citep[][appendix~B2]{moriche:20a}. 
Note that the present discussion equally applies to fluid-solid
coupling methods other than of immersed-boundary type (cf.\
\S\ref{sec:PRM-numerics-dlm}-\ref{sec:PRM-numerics-lbm}). 

Let us now turn to the choice for the number of Lagrangian marker
points to be distributed at the fluid-solid interface. It is common
practice to use a marker density comparable to the density of the
Eulerian grid points \citep{uhlmann:04}. 
It is also intuitively clear that the allowed minimum density is such
that the support of the discrete delta function at neighboring marker
points just overlaps -- otherwise the surface becomes permeable. 
\citet{zhou:21} have considered the volume $\Delta V_l$ in
\eqref{eq:PRM_ibm_spread_1} as an ajustable weight factor, and
their analysis then somewhat surprisingly shows that the optimal
choice (from the point of view of efficiency) corresponds to the
minimum marker density. This is due to the fact that the amplitude of
the velocity error (under the above mentioned condition on overlapping
delta functions, and for sufficiently small time steps) does not
significantly affect the value of the immersed-boundary force
\cite{zhou:21}.
We will return to this point in
\S~\ref{sec:PRM-numerics-ibm-refinements}.  
\subsection{%
  \revision{Which}{What level of}
  accuracy can be achieved with the immersed boundary
  method?
}
\label{sec:PRM-numerics-ibm-accuracy}
\citet{beyer:92} have shown for one-dimensional problems with singular
source terms that the immersed boundary method with discretized delta
functions commonly employed \citep[e.g.][]{roma:99} is only first-order
accurate in space when the error is measured on the Eulerian grid.
This convergence result also applies to the force acting on the
submerged body, as has been demonstrated by \citet{zhou:21}. 

\begin{figure}
  \begin{minipage}{1ex}
    $s$
  \end{minipage}
  \begin{minipage}{.45\linewidth}
    \centerline{$(a)$}
    \includegraphics[width=\linewidth]
    {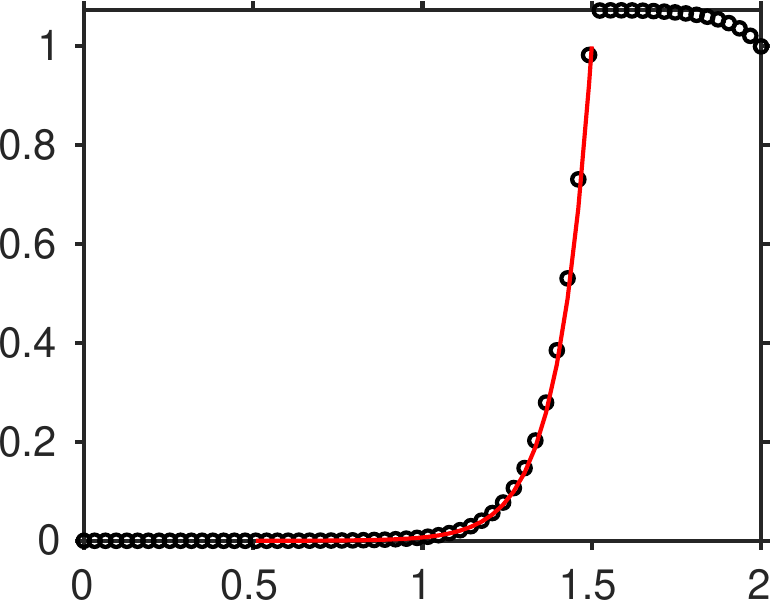}
    \hspace*{-.88\linewidth}\raisebox{.5\linewidth}{%
      \begin{minipage}{1ex}
        \rotatebox{90}{error}
      \end{minipage}
      \begin{minipage}{.5\linewidth}
        \includegraphics[width=\linewidth]
        {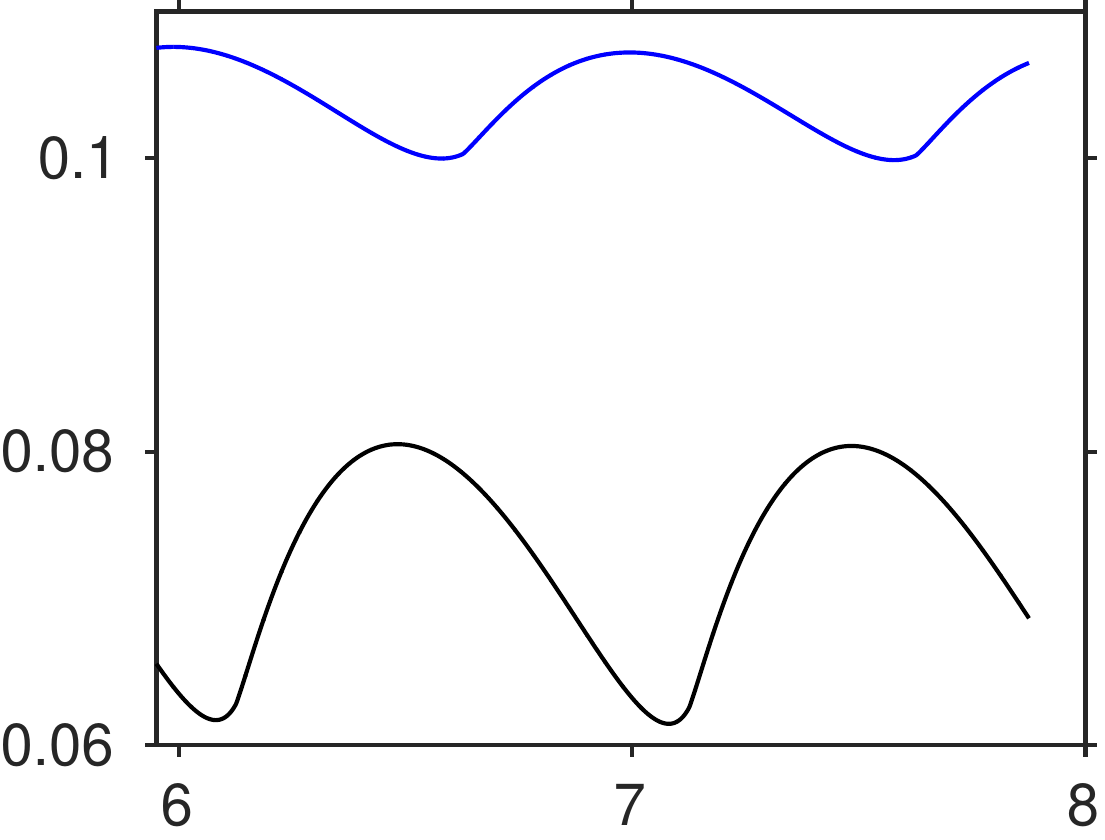}
        \\
        \centerline{$t/T_\Delta$}
      \end{minipage}
    }
    \\
    \centerline{$x/L$}
  \end{minipage}
  \hfill
  \begin{minipage}{1ex}
    \rotatebox{90}{error}
  \end{minipage}
  \begin{minipage}{.45\linewidth}
    \centerline{$(b)$}
    \includegraphics[width=\linewidth]
    {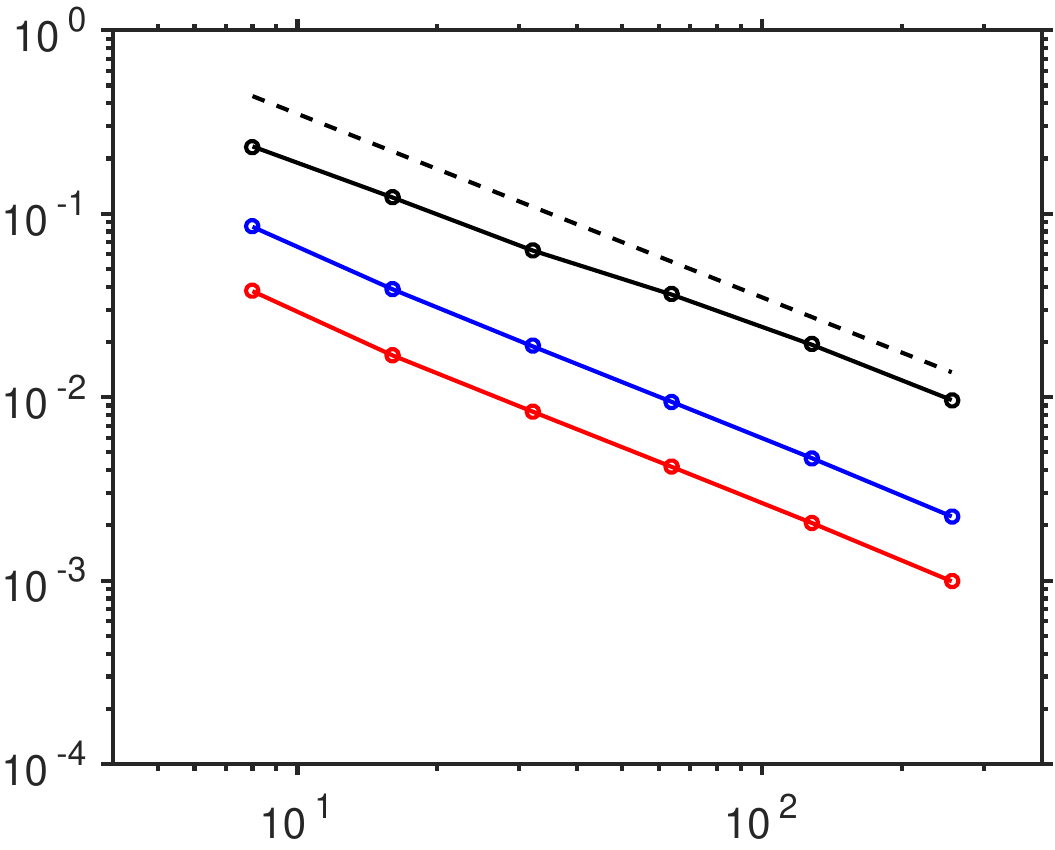}
    \\
    \centerline{$L/\Delta x$}
  \end{minipage}
  \caption{%
    Scalar advection-diffusion problem in one space dimension
    \eqref{eq:PRM_ibm_adv_diff1}, computed with the immersed boundary
    method in a standard second-order finite-difference discretization
    on a uniform grid and discretized in time with an explicit Euler
    scheme.  
    $(a)$ Solution for $Pe=10$, $u_w/u_{adv}=0.1$ at time
    $t/T_{adv}=2.5$ (where $T_{adv}=L/u_{adv}$ and $u_{adv}=c-u_w$),
    computed with
    $L_x/L=2$, $x_1(t=0)/L=0.25$, $L/\Delta x=31.5$, 
    $\sigma=c\Delta t/\Delta x=0.0087$, $\beta=D_s\Delta t/\Delta x^2=0.025$.
    The solid red line corresponds to the exact solution in $I_p$.
    The inset shows the relative error of the scalar boundary flux,
    $D_s s_{,x}(x_2)$, as a function of time, where $T_\Delta=\Delta
    x/u_w$ is the interval during which the immersed boundary
    traverses one grid cell (the black curve is for the three-point
    delta function of \citep{roma:99}, the blue curve for the
    four-point function of \citep{peskin:72}). 
    $(b)$ Convergence of the solution under spatial refinement. The
    lines with symbols indicate different error norms (L$_\infty$,
    {\color{blue}L$_1$,} {\color{red}L$_2$}) of the scalar field
    evaluated on the Eulerian grid points $x_i\in I_p$, 
    while the dashed line is proportional to $\Delta x/L$.
    The time step was kept constant, such that $\sigma=0.0011$ and
    $\beta=0.025$ on the finest grid. 
  }\label{fig:PRM_advdiff1}
\end{figure}
Let us consider a simple one-dimensional problem which will illustrate
some of the features of the immersed-boundary method. We will solve
the advection-diffusion equation for a scalar quantity $s(x,t)$, viz.
\begin{equation}\label{eq:PRM_ibm_adv_diff1}
  s_{,t}+c\,s_{,x}=D_s\,s_{,xx}
  \,,
\end{equation}
\revision{where $c>0=cst.$ is the advection speed}{%
  where $c>0$ is the constant advection speed}
and $D_s$ is the diffusivity.
The spatial interval of interest is $I_p=[x_1,x_2]$ with
$x_2=x_1+L$. The problem with Dirichlet boundary conditions
$s(x_1)=s_1$ and $s(x_2)=s_2$ has a steady-state solution
\begin{equation}\label{eq:PRM_ibm_advdiff_sol}
  s=s_1+(s_2-s_1)\,(\exp(Pe \,\tilde{x}/L)-1)/(\exp(Pe)-1)
  \,
\end{equation}
where $Pe=(c-u_w)L/D_s$ is the Peclet number, $u_w=cst.$ is the
velocity of the immersed boundaries, and $\tilde{x}=x-u_w\,t$ is the
position in a frame of reference moving with $u_w$. 
The asymptotic solution, which features an exponential boundary layer
that steepens with $Pe$ (figure~\ref{fig:PRM_advdiff1}$a$), shall be
computed on a fixed grid in the interval $I_c=[0,L_x]$, using the
immersed boundary method to enforce the conditions on $s$ at $x_1$,
$x_2$, while imposing the same Dirichlet conditions at the outer
domain boundaries (i.e.\ $s(0)=s_1$, $s(L_x)=s_2$) in a conventional way. 
We choose the computational domain such that $0<x_1$ and
$x_2<L_x$ for the duration of the computation (alternatively, one can
also impose periodic outer boundary conditions).

The results summarized in figure~\ref{fig:PRM_advdiff1} show that
the solution computed with the immersed boundary method is
well-behaved, and that it is indeed first-order accurate in space when
measured (using different norms) on the Eulerian grid points for which
$x_i\in I_p$.
The same is true for the scalar boundary flux (equivalent to a Nusselt
or Sherwood number, not shown) which would be the typical quantity of
interest here. Note that the flux is systematically overestimated, a
result which carries over to the drag force when applying the immersed
boundary method to the Navier-Stokes equations
\citep[cf.\ the discussion in][]{zhou:21}. For this reason some
authors have proposed to retract the Lagrangian force points from the
actual particle surface based on empirical corrections
\citep{breugem:12,luo:19}.
\revision{}{%
  Other authors have proposed immersed boundary schemes which avoid to 
  spread the forcing term into the volume occupied by the fluid phase
  \citep{tennetti:11}, thereby maintaining the momentum balance
  untainted.  
}
It is also important to note in the present
figure~\ref{fig:PRM_advdiff1}$a$ (inset) that the amplitude of the 
error varies smoothly while the forcing points ($x_1$, $x_2$) are
translating across the Eulerian grid.
A number of kernel functions for the interpolation
\eqref{eq:PRM_ibm_interp_1} and spreading \eqref{eq:PRM_ibm_spread_1} 
of quantities between Eulerian and Lagrangian positions
\revision{has}{have} 
been proposed
in the literature, and the reader is referred to \citep{peskin:02}
for a thorough discussion of the invariance properties and moment
conditions which can be achieved. Due to its support of only three
grid points in each linear dimension the variant stated in
\citep{roma:99} has been widely used for PR-DNS.
Discretized delta functions with broader support generally yield 
smoother signals during grid traverse; however, they typically lead to
larger errors in the predicted fluxes (cf.\
figure~\ref{fig:PRM_advdiff1}$a$, inset).

\revision{}{%
  It should be noted that second-order spatial accuracy can be
  achieved by modifying the finite-difference operators in the
  vicinity of the immersed boundary
  \citep[e.g.][]{mark:08,mittal:08}. 
  These ``sharp-interface'' or ``ghost-cell'' methods, however, are
  significantly more complex from the computational point of view, 
  especially when systems with a large number of mobile particles are
  considered.  
}
\subsection{%
  Further refinements}
\label{sec:PRM-numerics-ibm-refinements}
It appears reasonable to ask how well the immersed boundary method
manages to impose the desired velocity at the Lagrangian marker
points at the end of a given time step. \citet{gsell:21} have shown
that the discrepancy 
\revision{}{%
  (i.e.\ the difference between the updated and the desired velocities
  at the Lagrangian marker points)} 
depends essentially on the reciprocity of the 
interpolation and spreading operators
(\ref{eq:PRM_ibm_interp_1},\ref{eq:PRM_ibm_spread_1}).
When considering an isolated Lagrangian force point (such as on either
boundary in the one-dimensional equation \ref{eq:PRM_ibm_adv_diff1}),
the discrepancy is proportional to the deviation from unity of the 
sum (over its support) of the square of the discrete delta function. 
In order to reduce the Lagrangian velocity error, \citet{gsell:21}
propose to multiply the Lagrangian force relation
\eqref{eq:PRM_ibm_fibm_lag_1} by a constant correction factor which
only depends on the choice of the kernel function. In the case of an
isolated Lagrangian marker point and for a scalar equation, this
procedure guarantees that the desired velocity is exactly enforced. In
the case of multi-dimensional immersed interfaces (e.g.\ a sphere),
the overlap between the supports of delta functions centered at
neighboring Lagrangian points makes this procedure approximate. 
It should be noted, however, that the reduction (or even elimination)
of the velocity error at the Lagrangian marker points does not
guarantee a similar reduction of the error at the Eulerian grid points
in the fluid domain.
In fact, the error shown in
figure~\ref{fig:PRM_advdiff1}$(b)$ remains practically identically 
when the correction proposed in \cite{gsell:21} is applied, and the
same is true for the interface flux (not shown). 
This is consistent with the analysis of \citet{zhou:21} which we have
already discussed above.
Further investigations involving freely-moving particles should be
performed in the future in order to explore the potential of this
approach. 

Incidentally, the so-called ``multi-direct'' forcing approach 
\citep{luo:07}, where the immersed boundary force computation
(\ref{eq:PRM_ibm_interp_1}-\ref{eq:PRM_ibm_spread_1})
is repeatedly carried out per time step while the force field
$\mathbf{f}^{n+1/2}$ is being summed up, can be interpreted as an
iterative way of achieving essentially the same effect (in the limit
of an infinite number of iterations) as through the explicit
correction method of \citet{gsell:21}.  

Another point to consider is the consequence of using the immersed
boundary method in the context of a fractional step method. In this
case the modification introduced by the projection step leads to a
discrepancy which typically scales with the time step and with the
viscosity. One remedy is, therefore, to employ relatively small time
steps. 
With efficiency in mind, it would be preferable to construct
interpolation and spreading kernels which are divergence-free in a
discrete sense.  
\citet{bao:17} have proposed such an ansatz, which, however, requires
the solution of additional global Poisson problems at each time step.
Alternatively, \citet{taira:07} have shown how to absorb an  
immersed-boundary-like procedure into the projection step of a fractional
step method with the aid of an algebraic procedure that is somewhat
related to the Lagrange multiplier method (cf.\
\S~\ref{sec:PRM-numerics-dlm}). 
However, it is not clear how their technique
can be applied to freely-mobile 
immersed bodies. 
\input{chap1_particle_motion}

%% file: chap1_particle_motion.tex
\subsection{%
  Describing the motion of rigid particles}
\label{sec:PRM-numerics-newton-euler}
In the PR-DNS approach the Newton-Euler equations for the motion of
rigid bodies
(1.5)-(1.6)
need to be solved for each particle, with the hydrodynamic
force and torque
(1.7)-(1.8) 
being
directly obtained from the numerical approximation of the
Navier-Stokes equations at sufficient accuracy.
Since a direct evaluation of the surface integrals in
(1.7)-(1.8) 
is relatively costly, it is computationally more efficient to express
the hydrodynamic force and torque indirectly with the aid of the
immersed-boundary forcing term
which is already available.
More specifically, it can be shown that
under the assumption of rigid body motion
($\mathbf{u}=\mathbf{V}_p+\boldsymbol{\Omega}_p\times\mathbf{r}$,
where $\mathbf{r}$ is the distance from the center of mass) 
throughout the volume occupied by the particle we have:
\begin{equation}\label{eq:PRM_velocity-rate-of-change-int-rigid-body}
  \frac{d}{dt}\oint_{\partial\mathcal{V}_s^{(i)}}\mathbf{u}\,dV
  =V_p\,\frac{d\mathbf{V}_p^ {(i)}}{dt}
  \,,
\end{equation}
and analogously for the moment of moments \citep{uhlmann:04}.
Note that in the present section it is understood that $\mathbf{u}$
(without any subscript for a phase) refers to the composite velocity
field defined throughout the joint volume occupied by either the fluid
or the particles. The assumption of rigid body motion inside the
volume occupied by the solid, therefore,
allows us to express the hydrodynamic force acting on the $i$th
particle as: 
\begin{equation}\label{eq:PRM_net-particle-hydro-force-1}
  \mathbf{F}_h^ {(i)}=
  \rho_fV_p\,\frac{d\mathbf{V}_p^ {(i)}}{dt}
  -\rho_f\sum_{l=1}^{N_L^ {(i)}}\mathbf{F}(\mathbf{X}_l^ {(i)})\Delta V_l^ {(i)}
  \,,
\end{equation}
(where we have added the particle index as super-scripts to the
Lagrangian quantities for notational consistency), and again
analogously for the torque. Relation
\eqref{eq:PRM_net-particle-hydro-force-1} expresses the fact that the 
net force arises from the sum of one contribution which is described by the
acceleration of the composite fluid (which is treated as a single
phase with density $\rho_f$) and another contribution due to the applied
immersed boundary forcing (defined in \ref{eq:PRM_ibm_fibm_lag_1}).
The advantage of using \eqref{eq:PRM_net-particle-hydro-force-1} is
that the hydrodynamic force can be determined at practically no
additional cost, which typically makes tracking particles very
efficient in the 
context of immersed boundary methods (except for very dense
suspensions, where the cost of treating solid-solid contact becomes
significant, cf.\ chapter~11).
On the downside, the use of relation
\eqref{eq:PRM_net-particle-hydro-force-1} leads to a singularity for
the density-matched case ($\rho_p=\rho_f$), as the particle
acceleration then drops out of equation (1.5). 
However, for the time being let us restrict our attention to
\revision{dense particles;}{%
  particles with density larger than the fluid density;} 
the density-matched case and the treatment of light
particles will be discussed below.

We are now in a position to put all ingredients together, namely:
(i) the method to determine the fluid-solid interface coupling term 
(equations~\ref{eq:PRM_ibm_momentum_2}-\ref{eq:PRM_ibm_spread_1}),
(ii) the hydrodynamic force \eqref{eq:PRM_net-particle-hydro-force-1},
and the analogous term for the torque, required to advance the
particle state variables (position and velocity),
(iii) the basic single-phase Navier-Stokes solver.
It remains to specify the type of inter-phase coupling to be
employed. Weak (or explicit) coupling,
\revision{where the most recent value of the remaining phase is used
  for the purpose of updating a given phase,}{%
  where the most recent values of the variables describing the state
  of one phase are used for the purpose of updating the other phase,} 
is the typical choice for the obvious reason of efficiency. In this
case a number of conditions for stable integration arise which
depend on the solid-fluid density ratio, on the details of the
temporal discretization, on the distribution of the Lagrangian
markers, and on the choice of the regularized delta function
\citep{zhou:21}. In practice, the standard weakly-coupled approach is
limited to particles with an excess density (\citet{uhlmann:04}
reports stable integration for $\rho_p/\rho_f>1.2$, while
\citet{zhou:21} have shown how this limit can be somewhat lowered by
adapting the value of the volume associated with the Lagrangian
points).

\citet{patankar:01} has proposed a method which allows to
treat the density-matched case ($\rho_p=\rho_f$) by computing the
rigid particle velocity through integration of the (computational)
velocity field over the particle volume, thereby circumventing the
above-mentioned singularity (cf.\ also \citep{sharma:05}) and
at the same time obviating the need to explicitly solve the
Newton-Euler equations
(1.5)-(1.6). 
%
Another interesting route has been taken by \citet{kim:06} who propose
to integrate the particle momentum equation with a Crank-Nicolson
method (i.e.\ with implicit coupling), and then show how the
evaluation of the hydrodynamic force at the new (unknown) time level
can be avoided through use of Lagrangian quantities.
As a result, the final formulation is non-iterative and free from a
singularity for the density-matched case. 
Applications of this method to the motion of light particles,
however, have not been published to our knowledge. 
\revision{}{%
  A somewhat different approach is taken by \citet{schwarz:15} who
  introduce a stabilizing term on both sides of the Newton-Euler
  equations for rigid particle motion (which is termed `virtual mass
  force'), allowing for small density ratios to be tackled.
  This technique has subsequently been refined for larger volume
  fractions \citep{tavanashad:20}. 
}
Yet another approach is due to \citet{tschisgale:17} (cf.\ also
\citep{kempe:12a}), who define an interface layer (with finite
thickness) to which the rigid-body assumption is restricted. This then
leads to a discrete particle momentum equation which is also free from
the above mentioned singularity, and which even remains valid in the
limit of vanishing particle mass. 

%% file: chap1_coupling_dlm.tex
\section{%
  Distributed Lagrange multiplier methods
}
\label{sec:PRM-numerics-dlm}
The Direct Forcing IB methods presented in \S~\ref{sec:PRM-numerics-ibm-direct-forcing} are powerful numerical methods. However, the forcing term is explicit in time and may create stability issues for low density ratio. The forcing term is also distributed on the particle surface only, hence leaving a non-physical spurious flow inside rigid particles. Implicit variants have been proposed in the form of a fixed-point iteration to compute the forcing term \cite{breugem:12,kempe:12a} while volume forcing has been recently suggested by \cite{moriche:20a}. However, the Distributed Lagrange Multiplier/Fictitious Domain, abbreviated DLM/FD in the literature and introduced by Glowinski and co-authors in 1999 \cite{Glowinski1999,glowinski1999b} naturally features implicit forcing and volume forcing. The foundation of the original DLM/FD
method is to derive a global variational formulation for the fluid and particle momentum conservation
equations that allows to solve the fluid momentum and mass
conservation in the entire fluid domain through the following symbolic operation:
\begin{equation}
\label{eqn:dlmfd_symbolic}
\begin{split}
&\int_{\Om\backslash \P}{(\mathrm{fluid\:\:momentum})\cdot\bv d\bx} + (\mathrm{particle \:\:translational\:\:momentum})\cdot\bV \\
&\qquad\qquad+ (\mathrm{particle\:\:angular\:\:momentum})\cdot\bxi = \bzero,    
\end{split}
\end{equation}
where $(\bv,\bV,\bxi)$ are the test functions for $(\bu,\bU,\bom)$. The fluid mass conservation is
easily extended from $\Om\backslash \P$ to $\Om$ as a rigid body
motion is mass conserving by construction. The extension of the fluid
momentum conservation equation from $\Om\backslash \P$ to $\Om$
requires a bit more care and a few rather simple technicalities that
the reader can straightforwardly find in \cite{Glowinski1999,glowinski1999b}. In summary, once extended in the
rigid particle region, the fluid is forced to satisfy a rigid body
motion through the introduction of distributed Lagrange multipliers
over both the particle surface and the particle volume. The resulting
set of DLM/FD equations with homogeneous Dirichlet boundary conditions
on $\partial \Om$ (without any loss of generality) written in a
variational form reads as follows: 
\begin{enumerate}
\item Combined momentum equations
\begin{equation} 
\label{eqn:dlmfd_fluid_mom}
\int_{\Om}{\rho_f\left(\DPP{\bu}{t}+(\bu\cdot\grad)\bu\right)\cdot\bv d\bx} -\int_{\Om}{p\Div{\bv} d\bx} +\int_{\Om}{\mu_f\grad{\bu}:\grad{\bv} d\bx} +\sum_{i=0}^{N-1}\int_{\P_i(t)}{\bl\cdot\bv d\bx}= \bzero,
\end{equation}
\begin{equation} 
\label{eqn:dlmfd_solid_mom}
\begin{split}
&\left(1-\frac{\rho_f}{\rho_p}\right)\left[ M_i\left(\DCP{\bU_i}{t}-\bg\right) \cdot\bV_i+\DCP{\bJ_i \bom_i}{t}\cdot\bxi_i\right]-\sum_{j}{\bF_{c,ij}\cdot\bV_i} -\sum_{w}\bF_{c,iw}\cdot\bV_i \\
&-\sum_{j}{\bF_{c,ij}\cdot\bxi_i\times\bR_{j}}-\sum_{w}\bF_{c,iw}\cdot\bxi_i\times\bR_{j}
-\int_{\P_i(t)}{\bl\cdot(\bV_i+\bxi_i\times \br) d\bx}=\bzero\:\:,\:\: i=0,\dots,N-1
\end{split}
\end{equation}
\begin{equation} 
\label{eqn:dlmfd_rbm}
\int_{\P_i(t)}{\bal\cdot(\bu-(\bU_i+\bom_i\times \br) d\bx}=0\:\:,\:\: i=0,\dots,N-1. 
\end{equation}

\item Continuity equation
\begin{equation} 
\label{eqn:dlmfd_mass}
-\int_{\Om}{q\Div{\bu} d\bx}=0.
\end{equation}
\end{enumerate}   
where $\bl$ denotes the distributed Lagrange multiplier vector related
to the rigid body constraint $\bu=\bU+\bom\times \br$, $\bF_{c,ij}$
and $\bF_{c,iw}$ denote the particle-particle and particle-wall
contact forces, respectively,  $(\bal,\bV)$ are the test
functions for $(p,\bl)$, respectively, $\bV_i$ and
$\bxi_i$ are simple functions the components of which are uniformly
equal to $1$ over $\P_i(t)$ and all functions belong to the
appropriate functional spaces \cite{Glowinski1999,glowinski1999b}. The factor $(1-\rho_f/\rho_p)$ is the result of extending the fluid momentum equation from $\Om\backslash \P$ to $\Om$ \cite{Glowinski1999}. Note that the formulation breaks down when $\rho_f\leq\rho_p$. i.e., for both neutrally buoyant and
  lighter particles. This issue can be fixed in various manners: (i)
  through the derivation of an alternative DLM/FD formulation
  \cite{Patankar2000} in which $\bD(\bu)=0$ or equivalently
  $\Div{\bD(\bu)}=0$ is forced inside $P(t)$, or (ii) through the
  explicit treatment of $-\frac{\rho_f}{\rho_p}\DCP{\bU_i}{t}$ and
  $-\frac{\rho_f}{\rho_p}\DCP{\bJ_i \bom_i}{t}$
  \cite{rahmani2014,seyed2019dynamics}. The variational formulation
  \refeq{eqn:dlmfd_fluid_mom}-\refeq{eqn:dlmfd_mass} naturally leads
  to a Finite Element spatial discretization
  \cite{Glowinski1999,glowinski1999b,Glowinski2000,Glowinski2001}, but
  this set of equations can be re-written in a non-variational form
  such that other spatial discretization schemes can be used such as
  Finite Difference \cite{yu2002,yu2004,Yu2006c,Yu2006heat,Yu2007b}
  and Finite Volume
  \cite{rahmani2014,wachs2015accuracy,esteghamatian2017particle,seyed2019dynamics}. 

\refeq{eqn:dlmfd_fluid_mom}-\refeq{eqn:dlmfd_mass} is a saddle-point problem corresponding to a constrained minimization problem. The first constraint is the usual fluid velocity divergence free, relaxed with the pressure $p$ as a Lagrange multiplier, while the second constraint is the rigid body motion inside particles, relaxed with the distributed Lagrange multiplier $\bl$. This problem is solved by an Uzawa algorithm. The expression of the hydrodynamic force and torque exerted on each particle can be derived from \refeq{eqn:dlmfd_solid_mom} by simple identification with a classical way of writing Newton's law for each particle, and reads as follows: 
\begin{align}
\label{eqn:fhydro_dlmfd}
\bF_{h,i} &= \int_{\P_i(t)}{\bl d\bx} + \frac{\rho_f}{\rho_p}M_i\DCP{\bU_i}{t}, \\
\bT_{h,i} &= \int_{\P_i(t)}{\br\times \bl d\bx} + \frac{\rho_f}{\rho_p}M_i\DCP{\bJ_i \bom_i}{t}
\end{align}
If we re-write $\sum_{i=0}^{N-1}\int_{\P_i(t)}{\bl\cdot\bv d\bx}=\bef$ in \refeq{eqn:dlmfd_fluid_mom}, the resulting equation is similar to an IB method with the two fundamental differences: 
\begin{enumerate}
\item the forcing term $\bef$ is implicit in the original DLM/FD formulation
 while it is explicit in the vast majority of IB formulations. 
\item the forcing term $\bef$ is active in the particle volume and on the
particle surface in DLM/FD methods while it is active on the
particle surface only in the vast majority of IB formulations. 
\end{enumerate} 
The implicit computation of the distributed Lagrange multiplier $\bl$ and consequently of the hydrodynamic force and torque equips the original DLM/FD formulation with strong stability properties, in particular as a function of the density ratio ${\rho_f}/{\rho_p}$. However, there is a corresponding additional computing overhead associated to the iterative process of the Uzawa algorithm or its preconditioned conjugate gradient variant. Note that in practice, \refeq{eqn:dlmfd_fluid_mom}-\refeq{eqn:dlmfd_mass} are never solved in a monolithic fashion (with the exception of \cite{Yiantsios2012}) but instead with a fractional step time algorithm \cite{Glowinski1999,yu2004,wachs2015accuracy}. To avoid the computing overhead related to the Uzawa algorithm, many authors have derived DLM/FD formulations with an explicit equation for the distributed Lagrange multiplier $\bl$ \cite{sharma:05,Veeramani2007,Yu2007direct,Gallier2014}. These direct-forcing DLM/FD formulations are indeed very close to the direct-forcing IB formulations discussed in \S~\ref{sec:PRM-numerics-ibm}.
The many variants of DLM/FD or closely related methods differ in
  the way the distributed Lagrange multiplier $\bl$ or the
  corresponding forcing term $\bef$ is discretized on the grid
  (Lagrangian versus Eulerian representation, see
  \cite{haeri:12,wachs:19}
  for an extended
  discussion). An important feature, beyond the accuracy of the
  spatial discretization that is at best second order, is whether this
  discretization is gradient discontinuity capturing at the particle
  surface or not. The use of Finite Element basis functions or IB-type
  regularized Dirac kernels to transfer quantities from the Eulerian
  grid to the Lagrangian grid and vice-versa implicitly involve a
  certain degree of smearing or smoothing accross the fluid-particle
  interface. Sharp representation of the fluid-particle interface are
  generally obtained with ghost node or cut-cell discretization
  schemes. In the context of DLM/FD methods, the implementation in
  \cite{wachs2015accuracy} is gradient discontinuity capturing and can
  be viewed as an implicit ghost node method with additional volume
  forcing. 

DLM/FD methods have been successfully applied to a large range of
particle-laden flow problems: low to high solid volume fraction,
viscous and inertial flow regimes, with spherical and non-spherical
particles, and with and without heat transfer. DLM/FD implementations
that represent particles on the Eulerian fluid grid by a set of points
need to homogeneously distribute these points on the particle surface
and in the particle volume. The particle surface distribution
algorithms are similar to those utilized in IB methods discussed in
\S~\ref{sec:PRM-numerics-ibm-lagpoints}, adding to the long list of
similarities between DLM/FD and IB 
\cite{uhlmann:04,wachs2015accuracy}. As expected, the spatial accuracy of the DLM/FD method is found to be on par with that of IB methods and approximately the same number of points per particle diameter is required to yield a given level of accuracy \cite{uhlmann:04,yu2004,Yu2007direct,wachs2015accuracy}. The reader further interested in DLM/FD methods is referred to the following recent review papers \cite{haeri:12,wachs:19}.

%% file: chap1_coupling_mesoscopic.tex
\section{%
  \revision{Lattice-Boltzmann-based methods}{%
    Boltzmann-equation based mesoscopic methods} } 
\label{sec:PRM-numerics-lbm}

\revision{}{As alternative computational fluid dynamics (CFD) methods, kinetics methods based on the Boltzmann equation
have been developed over the last 30 years. Here we focus mainly on the kinetic methods that can solve the 
hydrodynamic equations (the Navier-Stokes and Fourier energy equations) for continuum fluids ({\itshape i.e.}, the low Knudsen number limit). The available
kinetic methods of this type include the lattice Boltzmann method
(LBM)~\citep{chen1998},
the lattice Boltzmann flux solver (LBFS)~\citep{shu2014development-prm},
the gas kinetic scheme (GKS)~\citep{xu2001gas-prm}, the discrete unified gas kinetic scheme (DUGKS)~\citep{guo2013discrete-prm, guo2015discrete-prm}. In this section, we briefly but rigorously introduce LBM and DUGKS, with LBM being
an on-grid finite-difference implementation, and DUGKS being an off-grid finite-volume formulation.  
Their advantages and drawbacks will be briefly compared and discussed in the end.}

\subsection{ \revision{From Boltzmann to lattice-Boltzmann to
    Navier-Stokes}
  {From Boltzmann equation to Navier-Stokes-Fourier}}
\label{sec:PRM-numerics-lbm-basics}
Fluids consist of atoms, often arranged in molecules. In a classical molecular dynamics description of a gas or liquid we follow individual molecules by solving their Newtonian equations of motion that include interaction forces between the molecules ~\cite{allen1987,frenkel2002}. In a continuum approach the discrete (molecular) nature of matter is abandoned to make place for a description in terms of functions describing properties (pressure, density, velocity, temperature) that continuously depend on location in space and on time. Kinetic theory is a description in between the molecular (\textit{microscopic}) and the continuum (\textit{macroscopic}) description, and is usually called \textit{mesoscopic}.

The starting point of the mesoscopic description is Boltzmann’s concept of the distribution function $f\left(\mathbf{x},\boldsymbol{\xi },t\right)$ with $\mathbf{x}$ location, $\boldsymbol{\xi }$ velocity, and \textit{t} time. It represents the density of molecules in physical space as well as in velocity space. The SI unit of the distribution function is $\left[f\right]=\mathrm{kg}\cdot \frac{1}{\mathrm{m}^{3}}\cdot \frac{1}{\left(\mathrm{m}/\mathrm{s}\right)^{3}}$.

The total time derivative of the distribution function is 
\begin{equation}
\frac{df}{dt}=\frac{\partial f}{\partial t}+\frac{\partial f}{\partial x_{\beta }}\frac{dx_{\beta }}{dt}+\frac{\partial f}{\partial \xi _{\beta }}\frac{d\xi _{\beta }}{dt},
\label{eq:PRM_lbm_total_deriv_eq}
\end{equation}
where we use index notation for the location ($x_{\beta }$) and velocity ($\xi _{\beta }$) vector components and the summation convention (we sum over repeated Greek indices). In \eqref{eq:PRM_lbm_total_deriv_eq}, $\frac{dx_{\beta }}{dt}=\xi _{\beta }$ and $\frac{d\xi _{\beta }}{dt}$ is acceleration per unit volume and therefore $\frac{d\xi _{\beta }}{dt}=\frac{F_{\beta }}{\rho }$ with $\rho $ density and $\mathbf{F}$ (external) force \revision{}{per unit volume}. Collisions between molecules are reasons for change of the distribution function since they re-distribute momentum over \revision{}{the} velocity space. Combining the above notions leads to the Boltzmann equation
\begin{equation}
\frac{\partial f}{\partial t}+\xi _{\beta }\frac{\partial f}{\partial x_{\beta }}+\frac{F_{\beta }}{\rho }\frac{\partial f}{\partial \xi _{\beta }}=\Omega \left(f\right),
\label{eq:PRM_lbm_Boltzmann}
\end{equation}
where $\Omega \left(f\right)$ is the collision operator. It depends on the distribution function, i.e.\ on the way mass is locally distributed over velocity space. The Boltzmann equations is an advection equation with forces and collisions as source terms. In a molecular collision, \revision{}{mass, momentum and kinetic energy} are conserved so that \revision{}{three basic} properties of the collision operator are
\begin{align}
\iiint \Omega \left(f\right)d^{3}\xi &=0, \\
\iiint \boldsymbol{\xi }\,\Omega \left(f\right)d^{3}\xi &=\mathbf{0}, \\
\revision{}{\iiint \frac{\xi^2}{2} \Omega \left(f\right)d^{3}\xi } &=0.
\end{align}
A lot of physics goes into the collision operator as it \textendash{} in principle \textendash{} involves a wide spectrum of collision scenarios and intermolecular forces. In this context it is worthwhile highlighting the equilibrium distribution. If a volume \revision{}{of} gas is left by itself for long enough time, kinetic theory states that as a result of collisions its distribution function evolves towards an equilibrium distribution, the Maxwell-Boltzmann distribution:
\begin{equation}
f^{eq}\left(\mathbf{x},\boldsymbol{\xi },t\right)=\rho \left(\frac{1}{ \revision{}{2\pi RT} }\right)^{3/2}e^{ \revision{}{-\left| \boldsymbol{c} \right|^{2}
/ \left( 2RT \right) } }
\label{eq:PRM_lbm_Maxwell_Boltzmann}
\end{equation}
\revision{}{where $\bm{c}\equiv \bm{\xi}-\bm{u}$ is the peculiar velocity, $ \boldsymbol{u}$ is the mean molecular or local hydrodynamic velocity, $T$ is temperature and $m$ is molar mass, and $R$ is the universal gas constant.  
}

\revision{}{In the limit of a very small Knudsen number ({\itshape i.e.}, the ratio of molecular mean free path to system length scale), the macroscopic equations governing the fluid continuum }
 can be derived from the Boltzmann equation \eqref{eq:PRM_lbm_Boltzmann}. A mass balance equation follows from integrating the Boltzmann equation over the velocity space; a momentum balance equation follows from multiplying the Boltzmann equation with velocity and then integrating over velocity space; \revision{}{and an energy balance equation is obtained by multiplying the Boltzmann equation with $\xi^2/2$ and then
 integrating over velocity space. An interesting insight from these derivations is the realization that viscous stress $\tau_{\alpha\beta}$ and heat flux $q_{\alpha}$ are the results of deviations of distribution functions from equilibrium, namely, 
 }
 \begin{align}
\tau_{\alpha\beta} &=  - \iiint   (\xi_\alpha - u_\alpha )  (\xi_\beta - u_\beta ) \left( f - f^{eq} \right)   d^{3}\xi,  \label{eq:PRM_stress} \\
q_{\alpha} & =   \iiint  \frac{\left| \boldsymbol{\xi } -  \boldsymbol{u }\right|^{2} }{2}  (\xi_\alpha - u_\alpha )   f   d^{3}\xi. \label{eq:PRM_heatflux}
\end{align}

\revision{}{When the Boltzmann equation is used as a numerical algorithm for continuum fluids,   a simplified collision operator,
known as the Bhatnagar, Gross \& Krook (BGK) collision operator ~\cite{bhatnagar1954}, is often assumed }
\begin{equation}
\Omega\left(f\right)=-\frac{1}{\tau }\left(f-f^{eq}\right),
\label{eq:PRM_lbm_bgk}
\end{equation}
\revision{}{namely, the collision process relaxes the distribution towards the equilibrium with a single relaxation time $\tau $. }

\revision{}{When $\tau$ is assumed a small parameter, equivalent to the assumption of very low Knudsen number,
one can approximate $f$ as
\begin{equation}
f = f^{eq} - \tau \left[  \frac{\partial f^{eq} }{\partial t}+\xi _{\beta }\frac{\partial f^{eq}}{\partial x_{\beta }}+\frac{F_{\beta }}{\rho }\frac{\partial f^{eq}}{\partial \xi _{\beta }}  \right] + {\cal O} (\tau^2),~~~ {\mbox{with}}~\frac{\partial f^{eq}}{\partial \xi _{\beta }} = - \frac{(\xi_\beta - u_\beta)}{RT} f^{eq} +  {\cal O} (\tau), 
\label{eq:PRM_CE}
\end{equation}
which is known as the Chapman-Enskog approximation. It is through this approximation that the external force term can be moved to the right-hand side of the Boltzmann equation and be treated similarly as the collision term.
Eq.~\eqref{eq:PRM_lbm_Boltzmann} together with Eqs.~\eqref{eq:PRM_lbm_bgk} and ~\eqref{eq:PRM_lbm_Maxwell_Boltzmann} defines
a model Boltzmann equation,
\begin{equation}
\frac{\partial f}{\partial t}+\xi _{\beta }\frac{\partial f}{\partial x_{\beta }} = -\frac{\left(f-f^{eq}\right)}{\tau }
+  \frac{F_{\beta} (\xi_\beta - u_\beta) }{\rho RT } f^{eq} \equiv \tilde{\Omega}.
\label{eq:PRM_model_boltzmann}
\end{equation}
Note that the collision term is modified to include the external force term, and the extended collision term is denoted by $\tilde{\Omega}$.
}

\revision{}{
Eq.~\eqref{eq:PRM_CE} provides the needed closure scheme for the viscous stress and the heat flux, up to the order ${\cal O}(\tau)$. 
It is straightforward to show that the macroscopic continuity, momentum, and energy equations are
 \begin{eqnarray}
 \frac{\partial \rho }{\partial t} +    \frac{\partial \left(\rho u_\beta \right) }{\partial x_\beta }  &=& 0, \label{eq:PRM_NSF1}\\
\frac{\partial }{\partial t}\left(\rho u_{\alpha }\right)+\frac{\partial }{\partial x_{\beta }}\left(\rho u_{\alpha }u_{\beta }\right) &=& -\frac{\partial p}{\partial x_{\alpha }}+  \frac{\partial  \sigma_{\alpha\beta} }{\partial x_{\beta }} + F_\alpha,  \label{eq:PRM_NSF2}\\
  &&~~ \hspace{-2cm} {\mbox{with}}~\sigma_{\alpha\beta} =
\mu \left(\frac{\partial u_{\alpha }}{\partial x_{\beta }}+\frac{\partial u_{\beta }}{\partial x_{\alpha }}-\frac{2}{3}\delta _{\alpha \beta }\frac{\partial u_{\gamma }}{\partial x_{\gamma }}\right)+ \mu^V \delta _{\alpha \beta }\frac{\partial u_{\gamma }}{\partial x_{\gamma }} , \label{eq:PRM_NSF3}\\
\frac{\partial }{\partial t}\left(\rho e \right)+\frac{\partial }{\partial x_{\beta }}\left(\rho u_{\beta } e \right) &=& - p \frac{\partial u_\beta }{\partial x_{\beta }}
- \frac{\partial}{\partial x_\beta} \left( q_\beta \right)  + \sigma_{\alpha \beta }   \frac{\partial u_\alpha }{\partial x_{\beta }} ,\label{eq:PRM_NSF4} \\
&&~~\hspace{-2cm} {\mbox{with}}~q_{\beta} = - k {\partial T \over \partial x_\beta}. \label{eq:PRM_NSF5}
\label{eq:PRM_lbm_nsf}
 \end{eqnarray}
 where the pressure is related to the temperature by the ideal gas equation of state $p=\rho RT$, the internal energy $e=C_v T$ with the specific heat at constant volume $C_v=3R/2$,  
 the specific heat at constant pressure $C_p=5R/2$,
 the dynamic shear viscosity $\mu = p \tau$, and the conductivity $k= C_p p\tau$.
 The value of bulk viscosity $\mu^V$ depends on whether $T$ is treated as a variable (the general compressible fluid) or is assumed to 
 be a constant ({\itshape i.e.}, so-called
 athermal model): for the former case  $\mu^V = 0$, and the latter case  $\mu^V = 2\mu/3$. 
 The simple model above yields a specific heat ratio of $5/3$ and a Prandtl number of unity. The value of specific heat ratio
 can be adjusted by introducing additional internal degrees of freedom~\citep{guo2015discrete-prm}, and the value of Prandtl number 
 can be adjusted in several ways, {\itshape e.g.}, by altering the equilibrium distribution or by adding a properly designed source term~\citep{chen2021inverse-prm}.
 }

\revision{}{%
What we have set up is an algorithm that uses a simplified Boltzmann equation to represent the Navier-Stokes-Fourier system, Eqs.~(\ref{eq:PRM_NSF1}) through~(\ref{eq:PRM_NSF5}). The essential idea of the mesoscopic methods is to numerically solve
Eq.~\eqref{eq:PRM_model_boltzmann} instead of the Navier-Stokes-Fourier system directly. Two methods of solving Eq.~\eqref{eq:PRM_model_boltzmann} are presented next.
}
\subsection{ \revision{}{The on-grid lattice-Boltzmann method}} 
\revision{}{The lattice-Boltzmann equation (LBE) is an efficient method to solve
 the model Boltzmann equation such as
Eq.~(\ref{eq:PRM_model_boltzmann}).} 
It involves a triple discretization of the model Boltzmann equation: space $\mathbf{x}$,  time $t$,  and particle velocity $\boldsymbol{\xi}$. 
The discrete velocity set of size $n+1$ is denoted as $\mathbf{e}_{\mathbf{i}}, i=0\ldots n$, the distribution function as $f_{i}\left(\mathbf{x},t\right)$ which is now understood as the density of molecules at location $\mathbf{x}$ and time $t$ having velocity $\mathbf{e}_{\mathbf{i}}$. %
\revision{}{Integrating Eq.~(\ref{eq:PRM_model_boltzmann}) along the path associated with $\mathbf{e}_{\mathbf{i}}$, 
from $t$ to $t+dt$, using the trapezoidal rule for the right hand side, one obtains
\begin{equation}
f_{i}\left(\mathbf{x}+\mathbf{e}_{\mathbf{i}}dt, t+dt \right) - f_{i}\left(\mathbf{x},t\right) =
\frac{dt}{2} \left[ {\tilde \Omega}_{i}\left(\mathbf{x},t\right) +{\tilde \Omega} \left(\mathbf{x}+\mathbf{e}_{\mathbf{i}}dt,t+dt\right) \right].
\label{eq:PRM_lbm_s1}
\end{equation}
To remove the time implicity, a linear transformation is introduced
\begin{equation}
{\tilde f}_i \left(\mathbf{x},t\right) = f_i\left(\mathbf{x},t\right) - \frac{dt}{2} {\tilde \Omega} \left(\mathbf{x},t\right),
\end{equation}
It is straightforward to show that Eq.~(\ref{eq:PRM_lbm_s1}) can then be written in an explicit form as
\begin{equation}
{\tilde f}_{i}\left(\mathbf{x}+\mathbf{e}_{\mathbf{i}}dt, t+dt \right) = {\tilde f}_{i} \left(\mathbf{x},t\right) 
 -\frac{\left( {\tilde f}_i -f^{eq}_i\right)}{\tau_L }
+ \left[ 1 - \frac{1}{2\tau_L} \right]  \frac{F_{\beta} (e_{i,\beta} - u_\beta) }{\rho RT } f_i^{eq} dt,
\label{eq:PRM_lbm_s2}
\end{equation}
where 
\begin{equation}
\tau_L = \frac{\tau}{dt} +  \frac{1}{2} =  \frac{\nu}{c_s^2 dt} +   \frac{1}{2},~~ {\mbox{or}} ~ \nu =\left( \tau_L -  \frac{1}{2} \right) c_s^2 dt,
\label{eq:PRM_lbm_s3}
\end{equation}
where $\nu$ is the fluid kinematic viscosity.
}

\revision{}{
An important step in deriving LBE is to convert the integral velocity moments into discrete summations over
$\mathbf{e}_{\mathbf{i}}$, a process known as the Gauss-Hermite quadratures. This amounts to a multiplication of a pre-factor involving
only $\mathbf{e}_{\mathbf{i}}$, namely, $g_i = w_i (2\pi RT)^{D/2} \exp[e_i^2/(2RT)]{\tilde f}$.
The linearity of Eq.~(\ref{eq:PRM_lbm_s2}) implies that $g_i$ is governed by
\begin{equation}
g_{i}\left(\mathbf{x}+\mathbf{e}_{\mathbf{i}}dt, t+dt \right) = g_{i} \left(\mathbf{x},t\right) 
 -\frac{\left( g_i \left(\mathbf{x},t\right)  -g_i^{eq} \left(\mathbf{x},t\right)  \right)}{\tau_L }
+ \left[ 1 - \frac{1}{2\tau_L} \right]  \frac{F_{\beta} (e_{i,\beta} - u_\beta) }{\rho RT } g_i^{eq}\left(\mathbf{x},t\right)  dt,
\label{eq:PRM_lbm_final}
\end{equation}
where, for isothermal flows, the Hermite expansion~\citep{shan2006kinetic-prm} of the equilibrium yield
\begin{equation}
g_{i}^{eq}=w_{i}\rho \left[1+\frac{u_{\alpha }e_{i\alpha }}{c_{s}^{2}}+\frac{\left(u_{\alpha }e_{i\alpha }\right)^{2}}{2c_{s}^{4}}-\frac{ u^2 }{2c_{s}^{2}}  +   \frac{\left(u_{\alpha }e_{i\alpha }\right)^{3}}{6c_{s}^{6}}  
-\frac{ \left(u_{\alpha }e_{i\alpha}\right)  u^2 }{2c_{s}^{4}} \right],
\label{eq:PRM_lbm_equilibrium_distribution}
\end{equation}
where $c_s^2 = RT$, $w_{i}$ a set of weighing factors that depend on the exact definition of the velocity set $\mathbf{e}_{\mathbf{i}}$ (or from the Gauss-Hermite quadrature requirements), and $c_{s}=\sqrt{RT}$. 
The terms of ${\cal O}(u^3)$ is included here to ensure that the Navier-Stokes equation is exactly recovered under the 
Chapman-Enskog approximation~\citep{shan2006kinetic-prm}. Eq.~(\ref{eq:PRM_lbm_final}) is the well-known LBE, here derived from the model Boltzmann equation with the external force term included as a source term.
}

\revision{}{
Under the Gauss-Hermite quadratures, the hydrodynamic moments for isothermal flow can be written as
\begin{eqnarray}\label{eq:PRM_lbm_rho}
\rho \left(\mathbf{x},t\right)&=&\sum _{i=0}^n g_{i}\left(\mathbf{x},t\right), \\
\label{eq:PRM_lbm_rho_u}
\rho \left(\mathbf{x},t\right)\mathbf{u}\left(\mathbf{x},t\right)&=&\sum _{i=0}^n\mathbf{e}_{\mathbf{i}} g_{i}\left(\mathbf{x},t\right)
+\frac{ \mathbf{F} dt} {2}. 
\end{eqnarray}
}
\revision{}{For computational efficiency, it is desirable to choose a lattice velocity model with a small $n$.}

\revision{}{The lattice Boltzmann method is an {\itshape on-grid} implementation method, meaning that 
the  modeled ``molecules''  travel from a lattice node to a neighbouring node,
or stay where they were if $\mathbf{e}_{\mathbf{i}}=\mathbf{0}$.   While this on-grid implementation makes the coding very simple and advection formally precise, it has two obvious disadvantages. 
First, the space lattice grid has to be geometrically simple ({\itshape e.g.}, hexagon and square cells in 2D). 
Second, the on-grid implementation limits the choices of the velocity set $\mathbf{e}_{\mathbf{i}}$ or alternatively the achievable quadrature order of velocity moments of  $g_{i}$; the latter is essential for the proper recovery of the hydrodynamic equations. For a square or cubic grid, $dx = e_{i1}dt$ and the fifth-order 
Gauss-Hermite quadrature would require $e_{i1} =\sqrt{3RT} $. For convenience, $dx$ is taken as the space unit
and $dt$ the time unit in the lattice Boltzmann method, leading to setting $c_s^2 = RT = 1/3$ in the standard LB scheme.
}

A time step in an LB algorithm consists of executing \eqref{eq:PRM_lbm_final} for every node of the lattice. For this we need the equilibrium distribution function in each node \eqref{eq:PRM_lbm_equilibrium_distribution} and therefore density and velocity in each node. For the latter two variables we use \eqref{eq:PRM_lbm_rho} and \eqref{eq:PRM_lbm_rho_u}.
For isothermal flow, it follows that 
\begin{equation}
p=\rho c_{s}^{2} .
\label{eq:PRM_lbm_pressure}
\end{equation}
Since equation \eqref{eq:PRM_lbm_pressure} implies that $\partial p/\partial \rho =c_{s}^{2}$, $c_{s}$ can be interpreted as the speed of sound which is finite in an LB scheme. This can be easily understood since a distribution function \textendash{} and therefore information \textendash{} only travels to a neighbouring node over one time step. An LB scheme therefore is a compressible scheme for which a Mach number can be defined:
\begin{equation}
\mathrm{M}\equiv \left| \mathbf{u}\right| /c_{s}.
\label{eq:PRM_lbm_mach}
\end{equation}

An often used velocity set in two dimensions is named D2Q9 (two
dimensions, nine velocities); an often used set in three dimensions is
D3Q19. %
For reference we give here
their definitions, i.e.\ the discrete set of velocities in Cartesian
coordinates and weighting factors. 

D2Q9: 
\begin{eqnarray}\nonumber
  \mathbf{e}_{\mathbf{0}}=\left(\begin{array}{l}
                                  0\\
                                  0
                               \end{array}\right),
  \,\mathbf{e}_{\mathbf{1}}=\left(\begin{array}{l}
                                    1\\
                                    0
                                \end{array}\right),\,
  \mathbf{e}_{\mathbf{2}}=\left(\begin{array}{l}
                                  -1\\
                                  0
                                \end{array}\right),\,
 \mathbf{e}_{\mathbf{3}}=\left(\begin{array}{l}
                                 0\\
                                 1
                               \end{array}\right),\
 \mathbf{e}_{\mathbf{4}}=\left(\begin{array}{l}
                                 0\\
                                 -1
                               \end{array}\right),\,
 \mathbf{e}_{\mathbf{5}}=\left(\begin{array}{l}
                                 1\\
                                 1
                               \end{array}\right),\\
 \mathbf{e}_{\mathbf{6}}=\left(\begin{array}{l}
                                 -1\\
                                 1
                               \end{array}\right),\,
 \mathbf{e}_{\mathbf{7}}=\left(\begin{array}{l}
                                 -1\\
                                 -1
                               \end{array}\right),\,
 \mathbf{e}_{\mathbf{8}}=\left(\begin{array}{l}
                                 1\\
                                 -1
                               \end{array}\right),
  \qquad
  \end{eqnarray}
  \begin{equation}
    w_{0}=4/9,w_{1-4}=1/9,w_{5-8}=1/36.
  \end{equation}
  D3Q19:
  \begin{eqnarray}\nonumber
    \mathbf{e}_{\mathbf{0}}=\left(\begin{array}{l}
                                    0\\
                                    0\\
                                    0
\end{array}\right),\,\mathbf{e}_{\mathbf{1}}=\left(\begin{array}{l}
                                                     1\\
                                                     0\\
                                                     0
\end{array}\right),\,\mathbf{e}_{\mathbf{2}}=\left(\begin{array}{l}
                                                     0\\
                                                     1\\
                                                     0
\end{array}\right),\,\mathbf{e}_{\mathbf{3}}=\left(\begin{array}{l}
                                                     -1\\
                                                     0\\
                                                     0
\end{array}\right),\,\mathbf{e}_{\mathbf{4}}=\left(\begin{array}{l}
                                                     0\\
                                                     -1\\
                                                     0
\end{array}\right),\,
\mathbf{e}_{\mathbf{5}}=\left(\begin{array}{l}
                                0\\
                                0\\
                                1
\end{array}\right),\\\nonumber
\mathbf{e}_{\mathbf{6}}=\left(\begin{array}{l}
                                0\\
                                0\\
                                -1
\end{array}\right),\,
\mathbf{e}_{\mathbf{7}}=\left(\begin{array}{l}
                                1\\
                                1\\
                                0
\end{array}\right),\,\mathbf{e}_{\mathbf{8}}=\left(\begin{array}{l}
                                                     -1\\
                                                     1\\
                                                     0
\end{array}\right),\,\mathbf{e}_{\mathbf{9}}=\left(\begin{array}{l}
                                                     -1\\
                                                     -1\\
                                                     0
\end{array}\right),\,\mathbf{e}_{\mathbf{10}}=\left(\begin{array}{l}
                                                      1\\
                                                      -1\\
                                                      0
\end{array}\right),\,
\mathbf{e}_{\mathbf{11}}=\left(\begin{array}{l}
                                 1\\
                                 0\\
                                 1
\end{array}\right),\\\nonumber
\mathbf{e}_{\mathbf{12}}=\left(\begin{array}{l}
                                 -1\\
                                 0\\
                                 1
\end{array}\right),\,
\mathbf{e}_{\mathbf{13}}=\left(\begin{array}{l}
                                 -1\\
                                 0\\
                                 -1
\end{array}\right),\,\mathbf{c}_{\mathbf{14}}=\left(\begin{array}{l}
                                                      1\\
                                                      0\\
                                                      -1
\end{array}\right),\,
\mathbf{e}_{\mathbf{15}}=\left(\begin{array}{l}
                                 0\\
                                 1\\
                                 1
\end{array}\right),\,\mathbf{e}_{\mathbf{16}}=\left(\begin{array}{l}
                                                      0\\
                                                      -1\\
                                                      1
\end{array}\right),\\
\mathbf{e}_{\mathbf{17}}=\left(\begin{array}{l}
                                 0\\
                                 -1\\
                                 -1
    \end{array}\right),\,
    \mathbf{e}_{\mathbf{18}}=\left(\begin{array}{l}
                                                      0\\
                                                      1\\
                                                      -1
                                   \end{array}\right),\qquad
      \end{eqnarray}
\begin{equation}
w_{0}=1/3,w_{1-6}=1/18,w_{7-18}=1/36.
\end{equation}

\revision{}{The {\itshape on-grid} implementation in the LB algorithm makes it difficult to design a lattice velocity model
which has a sufficient order to recover the hydrodynamic equations.  
Even for an isothermal flow, an important observation of the Chapman-Enskog approximation is that
the viscous stress would require the third-order moment of the equilibrium distribution
be computed correctly; this requires the lattice
velocity model to have at least a sixth-order Gauss-Hermite quadrature accuracy. 
Unfortunately, both D2Q9 and D3Q19 only have a fifth order
quadrature accuracy, then there is an error (proportional to $M^3$) in computing the viscous-stress moment. 
For this reason, the standard LB algorithm makes the assumption of low-Mach number or incompressible flow,}
thus we need to make sure that $\mathrm{M}\ll 1$ everywhere in the domain. With $c_{s}$ of order one (in lattice units $\Delta /\Delta t$), the $\mathrm{M}\ll 1$ condition then implies $\left| \mathbf{u}\right| =\left| \mathbf{u}\right| \Delta t/\Delta \ll 1$. The latter expression shows that absolute velocity in lattice units is equivalent to a Courant number. The LB scheme is explicit, see \eqref{eq:PRM_lbm_final}. The incompressibility constraint is therefore a much stronger constraint than the CFL stability condition $\left| \mathbf{u}\right| \Delta t/\Delta <1$ \cite{ferziger2002} that usually applies to explicit schemes. In practical terms one designs simulations such that the Mach number stays below 0.1 \cite{chen1998}.
\subsection{ \revision{}{The generalized off-grid DUGKS method} }
In DUGKS, we instead apply a finite-volume treatment to Eq.~\eqref{eq:PRM_model_boltzmann}. In this approach,
the discrete particle velocities  $\mathbf{e}_i$ are no longer coupled with the space lattice. This comes with two advantages: first, the choice
of the discrete velocity set $\{ \mathbf{e}_i \}$ can be better optimized to achieve the required Gauss-Hermite order of accuracy; second,
a non-uniform or unstructured mesh can be easily incorporated.  It should also be noted that the model 
Boltzmann equation, Eq.~\eqref{eq:PRM_model_boltzmann}, is a more general description of fluid motion which can be used to treat flows at arbitrary Knudsen numbers. For these reasons, DUGKS have been used to simulate a variety of flows including
 thermal compressible or non-continuum flows.

To this end, the distribution function is first generalized to include additional degrees of freedom, 
written as $f\left(\bm{x,\xi,\eta,\zeta},t\right)$, where $\bm{x}=\left(x_1,{...},x_D\right)$  and $\bm{\xi}=\left(\xi_1,...,\xi_D\right)$, 
$D$ is the spatial dimension of the hydrodynamic velocity $u_i$, $\bm{\eta}$ denotes the remaining $(3-D)$ space of
the microscopic velocity,  an internal kinetic variable,  $\bm{\zeta}$, of dimension $K$ is introduced in order to
adjust the specific heat ratio of the model. Second, a source term $S_f$ is included to allow for arbitrary Prandtl number and 
different ratios of bulk to shear viscosity, as well as a source term in the macroscopic energy equation. Then 
Eq.~\eqref{eq:PRM_model_boltzmann} and Eq.~\eqref{eq:PRM_lbm_Maxwell_Boltzmann} become
\begin{equation}
\frac{\partial f}{\partial t}+\xi _{\beta }\frac{\partial f}{\partial x_{\beta }} = -\frac{\left(f-f^{eq}\right)}{\tau }
+  \frac{F_{\beta} (\xi_\beta - u_\beta) }{\rho RT } f^{eq} + S_f.
\label{eq:PRM_model_boltzmann2}
\end{equation}
\begin{equation}
    f^{eq}=\frac{\rho}{{(2\pi RT)}^{(3+K)/2}}\exp\left(-\frac{c^2+\eta^2+\zeta^2}{2RT}\right),
    \label{eq:ch-prm-Maxwell2}
\end{equation}
The internal energy $e$ per unit mass and the heat flux are now formulated by
\begin{equation}
    \rho e=\int {\frac{ \left( c^2+\eta^2+\zeta^2 \right) }{2}} f d\bm{\xi}d\bm{\eta}d\bm{\zeta},~~~
    q_\beta=\frac{1}{2}\int c_\beta(c^2+\eta^2+\zeta^2)fd\bm{\xi}d\bm{\eta}d\bm{\zeta}.
        \label{eq:ch-prm-hydro1}
\end{equation}

For efficient numerical implementation, we first integrate over $(\bm{\eta},\bm{\zeta})$, 
and  introduce two reduced distributions $g(\bm{x},\bm{\xi},t)$ and $h(\bm{x},\bm{\xi},t)$~\cite{guo2015discrete-prm}: 
\begin{equation}
    g(\bm{x},\bm{\xi},t)=\int f(\bm{x},\bm{\xi},\bm{\eta},\bm{\zeta},t)d\bm{\eta}d\bm{\zeta},~~~
    h(\bm{x},\bm{\xi},t)=\int \left(\eta^2 + \zeta^2\right) f(\bm{x},\bm{\xi},\bm{\eta},\bm{\zeta},t)d\bm{\eta}d\bm{\zeta}.   
\end{equation}
Correspondingly, the governing equations for $g(\bm{x},\bm{\xi},t)$ and $h(\bm{x},\bm{\xi},t)$ can be obtained from Eq.~\eqref{eq:PRM_model_boltzmann2} as: 
\begin{subequations}
\begin{align}
    \frac{\partial g}{\partial t} + \bm{\xi}\cdot\nabla_{\bm{x}}g  & = -\frac{g-g^{eq}}{\tau} +  \frac{F_{\beta} (\xi_\beta - u_\beta) }{\rho RT } g^{eq} + S_g, 
    \label{eq:ch-prm-BGK-g}\\  
    \frac{\partial h}{\partial t} + \bm{\xi}\cdot\nabla_{\bm{x}}h  & = -\frac{h-h^{eq}}{\tau} + \frac{F_{\beta} (\xi_\beta - u_\beta) }{\rho RT } h^{eq} + S_h,
    \label{eq:ch-prm-BGK-h}
\end{align}
\label{eq:ch-prm-BGK-reduced}
\end{subequations}
where $S_g=\int S_f d\bm{\eta}d\bm{\zeta}$, $S_h=\int (\eta^2+\zeta^2)S_f d\bm{\eta}d\bm{\zeta}$, the reduced equilibriums $g^{eq}$ and $h^{eq}$ are 
\begin{subequations}
\begin{align}
    g^{eq}&=\int{f^{eq}d\bm{\eta}d\bm{\zeta}}=\frac{\rho}{{(2\pi RT)}^{D/2}}\exp\left[-\frac{\left(\bm{\xi}-\bm{u}\right)^2}{2RT}\right],\\
    h^{eq}&=\int \left(\eta^2 + \zeta^2\right) f^{eq}d\bm{\eta}d\bm{\zeta}=\left(K+3-D\right)RTg^{eq}.
\end{align}
\label{eq:ch-prm-BGK-equilibrium}
\end{subequations}

Using the  Chapman-Enskog approximation, it can be shown that the above system will reproduce the fully
compressible Navier-Stokes-Fourier system, Eqs.~(\ref{eq:PRM_NSF1}) - (\ref{eq:PRM_NSF5}), with a shear viscosity $\mu = p\tau$. The expressions for the
bulk viscosity $\mu^V$ and conductivity $k$ depend on the details of the source term $S_f$, as shown in~\citep{wen2021improved-prm,chen2021inverse-prm}.

Next, we describe the DUGKS algorithm.  Eqs.~\eqref{eq:ch-prm-BGK-g} and~\eqref{eq:ch-prm-BGK-h} 
which can be written as
\begin{equation}
    \frac{\partial \phi}{\partial t} + \bm{\xi}\cdot\nabla_{\bm{x}}\phi  = \tilde{\Omega}_{\phi} =  {\Omega}_{\phi}  + Q_\phi,
    \label{eq:ch-prm-evol1}
\end{equation}
where $\phi = g$ or $h$, $Q_\phi = \phi^{eq} \left[ F_{\beta} (\xi_\beta - u_\beta) \right]  / (\rho RT )$. 

As a finite-volume method, the first step is to decompose the computational domain into a set of control volumes. Eq.~(\ref{eq:ch-prm-evol1}) is integrated over a cell $V_j$ located at $\bm{x}_j$ from time $t_n$ to $t_{n+1}$. 
The time integration is done using the mid-point rule for the convection term and the trapezoidal rule for the collision term, then the evolution equation becomes
\begin{equation}
    \tilde{\phi}^{n+1}_{j} = \tilde{\phi}^{+,n}_{j} - \frac{\Delta t}{|V_j|} F^{n+\frac{1}{2}},~~~~F^{n+\frac{1}{2}}\equiv \int_{\partial V_j}\left(\bm{\xi}\cdot\bm{n}\right)\phi \left(\bm{\xi},\bm{x},t_{n+1/2}\right) d\bm{S},
    \label{eq:ch-prm-evol2}
\end{equation}
where the time implicity has been removed by introducing the following linear transformations
\begin{equation}
    \tilde{\phi} = \phi - \frac{\Delta t}{2}\bar{\Omega}_\phi,~~~~\tilde{\phi}^{+} = \phi + \frac{\Delta t}{2}\bar{\Omega}_\phi,
\end{equation}
and they are understood as cell-averaged values, $F^{n+\frac{1}{2}}$
is the microscopic flux across the cell interface at  $t_{n+1/2}
\equiv t_n + 0.5 \Delta t$, and $\bm{n}$ is the outward unit vector
normal to the cell interface located at $\bm{x}_{b}$.  

All hydrodynamic variables can be obtained from the auxiliary
distribution $\tilde{\phi}$ directly. For the evolution of
Eq.~(\ref{eq:ch-prm-evol2}), the key step is to evaluate the
microscopic flux $F^{n+\frac{1}{2}}$ at the cell interface at half
time step which requires the reconstruction of the original
distribution $\phi\left(\bm{\xi},\bm{x}_b,t_{n+1/2}\right)$. In order
to obtain that, we integrate the Boltzmann equation
Eq.~(\ref{eq:ch-prm-evol1}) for a half time step $s=\frac{\Delta
  t}{2}$, from $t_n$ to $t_n + 0.5 \Delta t$,  along the
characteristic line with the end point located at the cell interface
$\bm{x}_b$, again the trapezoidal rule is applied to the collision term: 
\begin{equation}
    \phi\left(\bm{\xi},\bm{x}_b,t_{n+1/2}\right) - \phi\left(\bm{\xi},\bm{x}_b-\bm{\xi}s,t_{n}\right) = \frac{s}{2}\left[\tilde{\Omega}_\phi\left(\bm{\xi},\bm{x}_b,t_{n+1/2}\right) + \tilde{\Omega}_\phi\left(\bm{\xi},\bm{x}_b-\bm{\xi}s,t_{n}\right)\right].
        \label{eq:ch-prm-evol3p}
\end{equation}
Once again to remove time implicity, another two auxiliary distributions $\bar{\phi}=\phi - (s/2) \tilde{\Omega}_\phi$ and $\bar{\phi}^+=\phi + (s/2) \tilde{\Omega}_\phi$ are introduced. And Taylor expansion around the node $\left(\bm{\xi},\bm{x}_b, t_n\right)$ is applied to $\bar{\phi}^{+}\left(\bm{\xi},\bm{x}_b-\bm{\xi}s,t_{n}\right)$: 
\begin{equation}
    \bar{\phi}\left(\bm{\xi},\bm{x}_b,t_{n+1/2}\right) = \bar{\phi}^{+}\left(\bm{\xi},\bm{x}_b-\bm{\xi}s,t_{n}\right) = \bar{\phi}^{+}\left(\bm{\xi},\bm{x}_b, t_n\right) - \bm{\xi}s\cdot\bm{\sigma}_b, 
    \label{eq:ch-prm-evol3}
\end{equation}
where $\bm{\sigma}_b=\nabla_{\bm{x}}\bar{\phi}^{+}\left(\bm{\xi},\bm{x}_b, t_n\right)$. It is obvious that the right-hand side of the Eq.~(\ref{eq:ch-prm-evol3}) is explicit, $\bar{\phi}^{+}$ can be obtained by interpolation and the slope $\bm{\sigma}_b$ can be
estimated from the difference between the two neighboring cells for
smooth flows. Hydrodynamic variables at time $t_{n+1/2}$ can be
computed from the distribution function $\bar{\phi}\left(\bm{\xi},\bm{x}_b,t_{n+1/2}\right)$. 
Then the equilibrium $\phi^{eq}$ can be obtained and the original distribution can be computed from $\bar{\phi}$: 
\begin{equation}
    \phi\left(\bm{\xi},\bm{x}_b,t_{n+1/2}\right) = \frac{2\tau}{2\tau + s}\bar{\phi}\left(\bm{\xi},\bm{x}_b,t_{n+1/2}\right) + \frac{s}{2\tau+s}\phi^{eq}\left(\bm{\xi},\bm{x}_b,t_{n+1/2}\right) + \frac{\tau s}{2\tau + s} Q_{\phi}.
\end{equation}

With the original distribution $\phi\left(\bm{\xi},\bm{x}_b,t_{n+1/2}\right)$ computed at the cell interface $\bm{x}_b$ at time $t_{n+1/2}$, the microscopic flux $F^{n+\frac{1}{2}}$ can be obtained.
The distributions at $t_{n+1}$ can then be obtained from Eq.~(\ref{eq:ch-prm-evol2}). 

In the numerical implementation, the equilibrium distribution  should be expanded to fourth order for
thermal flows, or third order for the isothermal flows~\citep{wen2021designing-prm}. The required order for Gauss quadrature is
thus eighth order for thermal flow and sixth order for isothermal flows, although this can be relaxed when
the flow Mach number is small. 

 In terms of the distributions for the discrete velocities, the hydrodynamic variables can be evaluated as 
\begin{subequations}
\label{eq:CE}
\begin{align}
\label{eq:hydro2a}
&   \rho =   \sum_{\alpha}  \tilde{g}_{\alpha}=  \sum_{\alpha}  \tilde{g}_{\alpha}^+ = \sum_{\alpha}  \overline{g}_{\alpha} = \sum_{\alpha}  \overline{g}_{\alpha}^+  ,   \\
\label{eq:hydro2b}
&  \rho  \bu = \sum_{\alpha} \xi_{\alpha} \tilde{g}_{\alpha}  + {dt \over 2} {\boldmath b}= \sum_{\alpha} \xi_{\alpha} 
\tilde{g}_{\alpha}^+ -  {dt \over 2} {\boldmath b}
 = \sum_{\alpha} \xi_{\alpha} \overline{g}_{\alpha}   + {s \over 2} {\boldmath b} =  \sum_{\alpha} \xi_{\alpha} \overline{g}_{\alpha}^+   -   {s \over 2} {\boldmath b} , \\
&   \rho \left(  C_V T + {u^2 \over 2} \right)   =    {1\over 2} \sum_{\alpha} \left( \xi^2_{\alpha} \tilde{g}_{\alpha}  + \tilde{h}_{\alpha} \right) + {\Delta t \over 2} \rho \mathbf{u}\cdot \mathbf{b} = {1\over 2} \sum_{\alpha} \left( \xi^2_{\alpha} \overline{g}_{\alpha}  + \overline{h}_{\alpha} \right) + {s \over 2} \rho \mathbf{u}\cdot \mathbf{b}.
\end{align}
\end{subequations}
Therefore, $\rho (\bx_b , t_n+s)$ and $ \bu (\bx , t_n+s)$  can be obtained from $\overline{g}_{\alpha}  (\bx_b, t_n + s) $, which
can be used to compute  $f_{\alpha}^{(eq)}  (\bx_b, t_n + s)$. 

When compared to the treatment of the nonlinear  advection term  in the
conventional  Navier-Stokes solvers, we observe two important differences: (1) the advection or related flux term in DUGKS is linear and
as such the time evolution from $t_n$ for $t_n+s$,
Eq.~(\ref{eq:ch-prm-evol3p}), is formally precise, (2) the
interpolation
here 
is over space only at a single time $t_n$ while the interpolation in the Navier-Stokes solvers
involves both space and time.
The coupled collision and streaming treatment in Eq.~(\ref{eq:ch-prm-evol3p}) is exact,
and the space-only interpolation in Eq.~(\ref{eq:ch-prm-evol3}) maintains a low numerical dissipation of the overall scheme.
A mathematical analysis of the numerical dissipation in
DUGKS~\citep{zhu2017performance-prm} shows that
the former 
depends on both
the mesh size and the time step size (or equivalently the CFL number),
but is smaller than
a version which uses 
a direct interpolation without
considering the collision term. 
\subsection{ \revision{}{Treating the moving solid-fluid interfaces} }
\label{sec:PRM-numerics-Boltzmann-SFI}
\revision{}{%
Both LBM and DUGKS have been applied to simulate suspension flows, turbulent flows, and
turbulent flows laden with finite-size particles.} The theoretical and practical groundwork for particle-resolved
suspension simulations through an LB approach was laid in two papers
by
\citet{ladd1994a,ladd1994b}.
The papers are not
only among the first in terms of \revision{}{approximately} resolved suspension
simulations, they also give a comprehensive account of the LB method
and the way it relates to Newtonian fluid dynamics and how moving
no-slip boundaries can be accounted for.  \revision{}{In Ladd's approach, 
the  location of the no-slip surface is halfway the link between two lattice nodes that is cut by the surface, due to
the simple (or half-way) bounce back implemented as
\begin{equation}
f_i ({\boldsymbol x}_f, t+dt) = f^*_{\bar i} ({\boldsymbol x}_f,t) + 2\rho w_i { {\boldsymbol e}_i \cdot {\boldsymbol u}_w \over c_s^2},
\label{eq:PRM_lbm_SBB}
\end{equation} 
where ${\boldsymbol x}_f$ denotes the location of a fluid node near the solid boundary, ${\boldsymbol e}_i$ denotes a direction coming from the wall, and ${\boldsymbol e}_{\bar i} = -  {\boldsymbol e}_i$,
$f^*$ denotes the right hand side of Eq.~\eqref{eq:PRM_lbm_final} or the value after collision,
${\boldsymbol u}_w$ is the local wall velocity.
This results in a stair-step boundary condition that approaches the
actual surface upon refining the lattice.} 
There is fluid inside the particle surfaces which is kept there for convenience. 
\revision{}{This internal fluid is
  often considered to be 
  decoupled from the external fluid. 
}
The convenience lies in the fact that this approach avoids the need
for creating fluid when a moving particle uncovers a node and
destroying fluid when it covers a node. Reference \cite{ladd1994b}
deals with a number of practicalities when applying the approach to
multiple spherical particles suspended in fluid.  

\begin{figure}
  \centering
  \includegraphics[width=.35\linewidth]
  {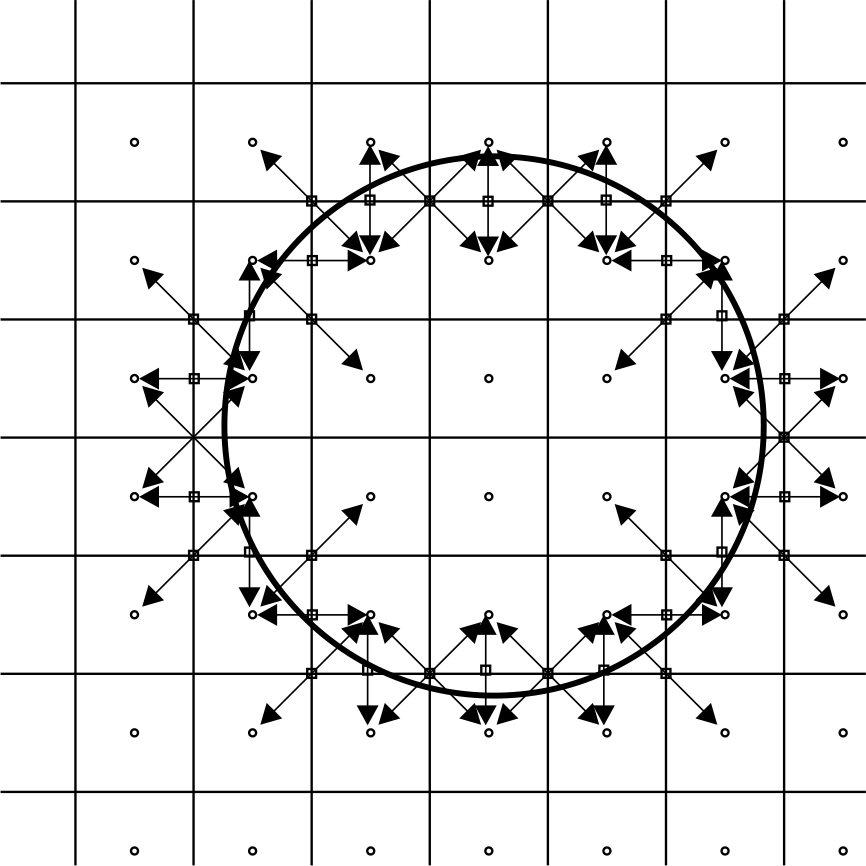}
  \hspace{1cm}
    \includegraphics[width=.55\linewidth]
  {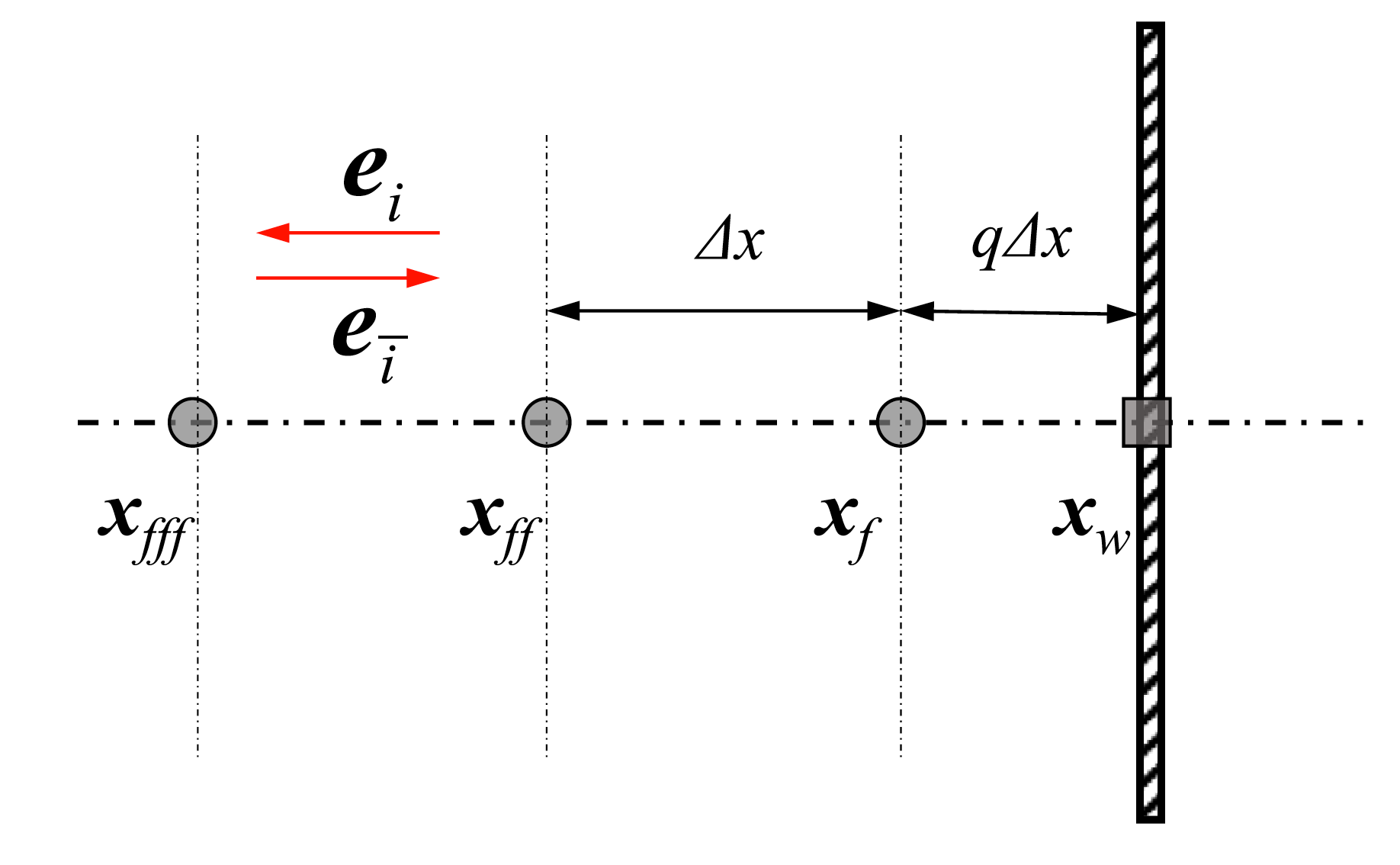}
  \caption{%
    Left panel: Half-way bounce-back representation of a circular no-slip boundary on a
    square lattice. Circular dots are lattice nodes sitting in the
    centre of square cells. The bounce-back conditions representing
    no-slip are imposed halfway the links between those two nodes that 
    intersect with the circle (square dots) \cite{ladd1994b}.
    Right panel: sketch used to illustrate the interpolated
    bounce-back scheme. 
 }
  \label{fig:PRM_lbm_sketch}
\end{figure}
\revision{}{The stair-step boundaries and associated particle shape fluctuations
when a particle moves over the lattice have motivated the interpolated bounce-back scheme or the use of
immersed boundaries in LB simulations.  It is well known that Eq.~\eqref{eq:PRM_lbm_SBB} is only of first-order accuracy for
a curved fluid-solid surface.   In order to ensure the second-order spatial accuracy in a bounce-back process,
a number of interpolated bounce-back (IBB) schemes have been introduced over the years,
with the conditional IBB scheme proposed by~\citet{bouzidi2001momentum-prm} and the unified IBB scheme by~\citet{yu2003viscous-prm} being
the most representative. The
location
of the solid wall is defined by $q=|{\boldsymbol x}_f - {\boldsymbol x}_w|/|{\boldsymbol x}_f - {\boldsymbol x}_b|$ (see Figure~\ref{fig:PRM_lbm_sketch} (b)).  In Bouzidi et al.’s scheme, when $q < 0.5$, a virtual distribution function is interpolated first so that the particle represented by this virtual distribution function ends precisely at ${\boldsymbol x}_f$ after the bounce back. On the other hand, when $q\ge 0.5$, the bounce back of distribution $f^*_{\bar i}$ is
performed first, and then interpolated back to ${\boldsymbol x}_f$. Namely,
 \begin{eqnarray}
f_i ({\boldsymbol x}_f, t+dt) &=& q(2q+1) f^*_{\bar i} ({\boldsymbol x}_f,t)  + (1+2q)(1-2q) f^*_{\bar i} ({\boldsymbol x}_{ff},t)  \nonumber \\
 &&  -q(1-2q)  f^*_{\bar i} ({\boldsymbol x}_{fff},t)   + 2\rho w_i { {\boldsymbol e}_i \cdot {\boldsymbol u}_w \over c_s^2},~~ q < 0.5,  \\
 f_i ({\boldsymbol x}_f, t+dt) &=& {1\over q(2q+1) } \left[ f^*_{\bar i} ({\boldsymbol x}_f,t) + 2\rho w_i { {\boldsymbol e}_i \cdot {\boldsymbol u}_w \over c_s^2}  \right]   + { 2q -1\over q } f_i ({\boldsymbol x}_{ff},t + dt) \nonumber \\
 && -   { 2q -1\over 1+ 2q } f_i ({\boldsymbol x}_{fff},t+dt)   ,~~ q \ge 0.5,  
\label{eq:PRM_Bouzidi}
 \end{eqnarray}
 Alternatively, Yu et al. designed a unified IBB scheme for all values of $q$.  First, a virtual distribution function is interpolated between 
 ${\boldsymbol x}_f$ and  ${\boldsymbol x}_{ff}$, which ends exactly at the wall location after streaming a grid spacing towards the wall. 
 Next, an instantaneous bounce-back happens right after the virtual distribution function arrives at the wall location. Finally, 
 the unknown distribution function is interpolated from this distribution and that at ${\boldsymbol x}_{ff}$. Together, this yields
 \begin{eqnarray}
 f_i ({\boldsymbol x}_f, t+dt) &=& 
 {q\over (2 + q)} f^*_{\bar i} ({\boldsymbol x}_f,t)  
 + { 2(1-q)\over (1 + q)} f^*_{\bar i} ({\boldsymbol x}_{ff},t)  
  - {(1 - q) q \over (1+q)(2+q)}  f^*_{\bar i} ({\boldsymbol x}_{fff},t)   \nonumber \\
 && \hspace{-2cm} + { 2q  \over 1+ q } f_i ({\boldsymbol x}_{ff},t + dt) 
  -   { q \over 2+ q } f_i ({\boldsymbol x}_{fff},t+dt)   
    + {4\over (1+q)(2+q)} \rho w_i { {\boldsymbol e}_i \cdot {\boldsymbol u}_w \over c_s^2} 
 \label{eq:PRM_Yu}
 \end{eqnarray}
In both the simple bounce-back and interpolated bounce-back schemes, the momentum exchange between the particle and the solid wall during bounce-back is well defined, thus by summing these exchanges, one can easily compute the hydrodynamic force and torque acting on
the solid particle.}
 
Initial attempts to apply the immersed boundary method in the context
of lattice-Boltzmann simulations were based on the feedback approach
due to 
\cite{goldstein:93}
and
alluded to in the present \S~\ref{sec:PRM-numerics-ibm}.
The approach is to locally apply forces on the fluid that oppose a difference between the fluid velocity at the immersed boundary and the desired velocity of the immersed boundary. Iteration or time stepping then establishes and maintains the velocity condition. An early \textendash{} 1999 \textendash{} application of this method is the representation of an impeller in a mixing tank in single-phase large-eddy simulations \cite{derksen1999}. The surface of the impeller is represented by a closely spaced set of off-lattice marker points with a typical nearest neighbour spacing of $0.5-0.7\Delta $. The fluid velocity at the marker points is determined through interpolation (linear or quadratic) from the lattice, as
in  Eq.~(\ref{eq:PRM_ibm_interp_1}).
\revision{}{%
We then update the force on the fluid responsible for imposing the velocity boundary condition at Lagrangian points $\mathbf{X}_l$ as
\begin{equation}
\mathbf{F}^{\left(n\right)} (\mathbf{X}_l) =\mathbf{F}^{\left(n-1\right)} (\mathbf{X}_l)  +\alpha 
\left( \mathbf{U}^{(d)}(\mathbf{X}_l)  -  \tilde{\mathbf{U}}_f(\mathbf{X}_l) \right),
\label{eq:PRM_lbm_ibforce}
\end{equation}
with $\mathbf{U}^{(d)}(\mathbf{X}_l)$ the velocity boundary condition
and $\alpha $ a relaxation parameter. The final step is to distribute
the forces at the Lagrangian points back to their Eulerian lattice
nodes, similar to Eq.~(\ref{eq:PRM_ibm_spread_1}).} 

The stability and accuracy of the method hinges on the free parameter
$\alpha $ \cite{derksen1999}. Having a free parameter, as noted in
\S~\ref{sec:PRM-numerics-ibm}, is a disadvantage. On the positive
side, the method largely retains the locality of the arithmetic
operations inherent to the LB method. This feature of the LB method
enables simple computational parallelization strategies.  

As noted, initially the above method was applied for imposing the
moving no-slip condition at the surface of a revolving
impeller. Extending it to solid particles moving through fluid is
straightforward \cite{tencate2004}. Different from an impeller,
however, the linear and angular motion of the particle and therefore
the solid surface velocity is not known a priori and needs to be
solved. This involves solving the equations of linear and rotational
motion of each particle.
\revision{his is the subject of chapter~11}{This is the
  subject of \S~\ref{sec:PRM-numerics-newton-euler}}. 
The equations of motion include the hydrodynamic force and torque on each particle. These are directly available through integration of the force distribution that is responsible for imposing the immersed boundary \eqref{eq:PRM_lbm_ibforce} over the surface of the particle. 

Alternative force-based immersed boundary methods in an LB context
have been proposed by e.g. \cite{feng2004}. They stay close to the
original ideas of
\citet{peskin:72}
(see also Section~\ref{sec:PRM-numerics-ibm}) of an ``elastic
restoration force'' that counteracts deformations of the immersed
solid object as a result of fluid flow. The restoring forces act on
off-grid marker points. They are distributed over neighbouring lattice
nodes through a regularized delta function \cite{feng2004}.  

\revision{The force-based immersed boundary methods discussed here have fluid
  inside a particle. As a consequence, the immersed boundary forces
  accelerate external as well as internal fluid. It is only the external
  force that has direct physical significance as the hydrodynamic force
  acting on the particle. As a result, one needs to distinguish between
  the external and internal force contributions. One way to do this is
  to assume that the fluid inside the particle moves as a solid
  body. This specifically is a good approximation when, in addition to
  immersed boundary marker points at the fluid-solid interface, a layer
  of marker points is added inside the particle \cite{derksen2007}.  
  Based on this assumption, the external force
  {$\mathbf{F}$} equals the overall immersed boundary force
  $\mathbf{F}_{\mathbf{IB}}$ minus the inertial force of the internal
  fluid \cite{derksen2007} 
  \begin{equation}
    \mathbf{F}=\mathbf{F}_{\mathbf{IB}}-\rho V_{p}d\mathbf{u}_{\mathbf{p}}/dt,
  \end{equation}
  with $\rho_f$ the fluid density and $V_{p}$ the volume of the
  particle. For the external torque an equivalent expression can be
  derived in terms of the overall immersed boundary torque and the
  rotational acceleration of the internal fluid.
  
  Given that $\mathbf{F}_{\mathbf{IB}}$ is a directly available output
  of the immersed boundary method, its inclusion in the equation of
  linear motion of the particles leads to inertia terms of the form
  $\left(\rho _{p}-\rho \right)V_{p}d\mathbf{u}_{\mathbf{p}}/dt$. For
  cases with particles and fluid having similar density the density
  difference and therefore the inertia term gets small which has
  problematic consequences for the stability of numerical integration of
  the equation of motion. Specific time stepping algorithms have been
  devised to mitigate these issues \cite{feng2009}.
}{%
  Due to the motion of virtual fluid inside the volume occupied by the
  solid particle in these force-based immersed boundary methods, care
  has to be taken when defining the hydrodynamic force/torque which
  enters the Newton-Euler equations for linear and angular particle
  acceleration. This is analogous to the discussion in
  \S~\ref{sec:PRM-numerics-newton-euler} leading to 
  equation~\ref{eq:PRM_net-particle-hydro-force-1}.
  Similarly as already discussed in
  \S~\ref{sec:PRM-numerics-newton-euler}, a singularity arises in the
  limit of density-matched particles. Several authors have proposed
  alternatives in the context of LBM methods \citep{feng2009}. 
}

\revision{}{%
  Returning back to the LB scheme, Eq.~\eqref{eq:PRM_lbm_final}, the last term used to implement the external  force
  is often simplified by inserting the Hermite expansion,  Eq.~\eqref{eq:PRM_lbm_equilibrium_distribution}, and keeping only terms up to
  ${\cal O}(Ma)$, namely,
  \begin{equation}
    \left( 1 - \frac{1}{2\tau_L} \right) \frac{F_{\beta} (e_{i,\beta} - u_\beta) }{\rho RT } g_i^{eq}\left(\mathbf{x},t\right)  dt 
    \approx   \left(1-\frac{1}{2\tau }\right)w_{i}\left(\frac{e_{i\alpha }}{c_{s}^{2}}+\frac{\left(e_{i\alpha }e_{i\beta }-c_{s}^{2}\delta _{\alpha \beta }\right)u_{\beta }}{c_{s}^{4}}\right)F_{\alpha } dt,
    \label{eq:PRM_lbm_si}
  \end{equation}
  which is the well-known Guo scheme~\citep{guo2002discrete-prm}.
}

\revision{}{The method to implement the boundary condition for a
  fixed solid wall in DUGKS is very similar to that in LBM, 
except that it is done at the cell interface
nodes~\cite{guo2013discrete-prm}.  
For moving solid-fluid interface, the immersed boundary method
described above is typically used, see, for example, the study
of~\citet{tao2018combined-prm} for particle-laden flows.  
}
\revision{}{%
\subsection{ {Remarks} }
\bgroup 
\def\arraystretch{1.0}%
\begin{table*}[!h]
\caption{Comparison of LBM and DUGKS.}
\hrule
\footnotesize
\begin{center}
\begin{tabular}
 { || r || p{4.5cm} | p{4.5cm} || }
         & {\bfseries LBM} &  {\bfseries DUGKS}    \\
\hline
Basic features & A finite-difference method along the characteristic
                 line   & A coupled finite-volume approach  \\ 
\hline
Discrete velocity set & More limited due to tight coupling with the
                        space grid & Uncoupled from the space grid,
                                     more flexibility \\ 
\hline
Mesh  & Typically uniform mesh & Nonuniform / unstructured mesh
                                 allowed \\ 
\hline
Coding  & Simple, only collision \& streaming & More involved due to
                                                the computation of
                                                interface fluxes \\ 
\hline
Numerical stability& Less stable & More stable  \\ 
\hline
Order of accuracy & Second order & Second order  \\
\hline
Numerical dissipation  & Minimal &  Slightly larger than LBM due to
                                   reconstructions at the cell
                                   interface \\ 
\hline
Comput. efficiency & High & Not as high as LBM  \\
\end{tabular} 
\end{center}
\label{tab:PRM-methods}
\hrule
\end{table*}
\egroup
{Before ending this section, several remarks are in order. First, we compare LBM and DUGKS in Table~\ref{tab:PRM-methods}.
  In general, they are comparable in terms of accuracy. LBM has some advantages over DUGKS in terms of computational efficiency, accuracy and coding
  simplicity, while DUGKS has been shown to be better in terms of numerical stability, mesh choice, and choice of particle velocity set.
  LBM is a relatively more mature approach, while DUGKS can be more easily extended to compressible and thermal flows.
}

Second, we compare briefly the interpolated bounce-back  (IBB) scheme IBB and the immersed boundary method (IBM) for handling moving
fluid-solid interface, in the context of mesoscopic methods. A direct
comparison was made recently by Peng et
al.~\cite{peng2019comparative-prm,peng2019motta-prm}.
Consistent with the observations and remarks in the present
\S~\ref{sec:PRM-numerics-ibm-accuracy}, 
these authors show that IBM  typically yields first-order accuracy locally when the regularized delta-function is employed to interpolate velocity from the Eulerian to Lagrangian mesh, and the resulting boundary force back to the Eulerian mesh. 
This first order in accuracy for IBM is observed for both the local velocity and hydrodynamic force/torque. Another problem of immersed boundary methods is that the local stress within the diffused fluid-solid interface tends to be significantly underestimated, which can 
significantly affect the accuracy of
computed viscous dissipation rate at the solid-fluid interfaces.
On the other hand, the interpolated bounce-back generally possesses a second- order accuracy for velocity, hydrodynamic force/torque, and local stress field. The main disadvantage of the interpolated bounce-back schemes is its higher level of fluctuations in the calculated hydrodynamic force/torque when a solid object moves across the grid lines.  
Nevertheless, \citet{peng2019motta-prm} show that, when a sufficient grid resolution is used, the two methods are able to provide accurate results for most of turbulent statistics in both the carrier and dispersed phases in turbulent particle-laden flow simulations. 

Finally, we wish to point out that handling accurately the hydrodynamic boundary conditions in mesoscopic methods remains an active
area of  research, particularly for moving solid-fluid interfaces in a turbulent flow. Issues such as Galilean invariance, pressure noises, and
thermal boundary conditions are not
fully resolved. While mesoscopic methods enjoy linear parallel scalability, the criterion for their numerical stability is not as well established as
the traditional NSF solvers.
}

%% file: chap1_reference_data.tex
\section{Reference data-sets} 
\label{sec:PRM-benchmark}
Since efficient algorithms for PR-DNS are non-boundary-conforming in
nature, their thorough validation is primordial. However, it is not
trivial to design  meaningful test cases which involve the full
non-linearity of the Navier-Stokes equations, coupled to rigid body
motion, suitable for convergence studies.
Even after a given algorithm has been fully validated in the general
sense, there always remains the task of calibrating the
spatio-temporal resolution to be employed in a given simulation.
For this purpose high-fidelity reference data 
involving a three-dimensional flow field is extremely useful. 

One way of generating such benchmark data is through
boundary-conforming simulations, preferably using a high-order
method. In this spirit the spectral-element method, with a
particle-attached grid and strong coupling between 
the fluid and solid motion, provides excellent reference data. In the
following we are discussing one such data-set in an unbounded system. 
Other high-fidelity data-sets featuring wall-bounded shear flows have
been described in \citep{zeng:05,zeng:07,zeng:09,zeng:10}. 
\subsection{Single particle settling in unbounded ambient fluid} 
\label{sec:PRM-benchmark-single-particle-settling-case}
\begin{figure} 
  \begin{minipage}{.13\linewidth}
    \centerline{$(a)$}
  \end{minipage}
  \begin{minipage}{.13\linewidth}
    \centerline{$(b)$}
  \end{minipage}
  \hfill
  \begin{minipage}{.3\linewidth}
    \centerline{$(c)$}
  \end{minipage}
  \begin{minipage}{.3\linewidth}
    \centerline{$(d)$}
  \end{minipage}
  \\
  \begin{minipage}{.13\linewidth}
    \includegraphics[width=\linewidth]
    {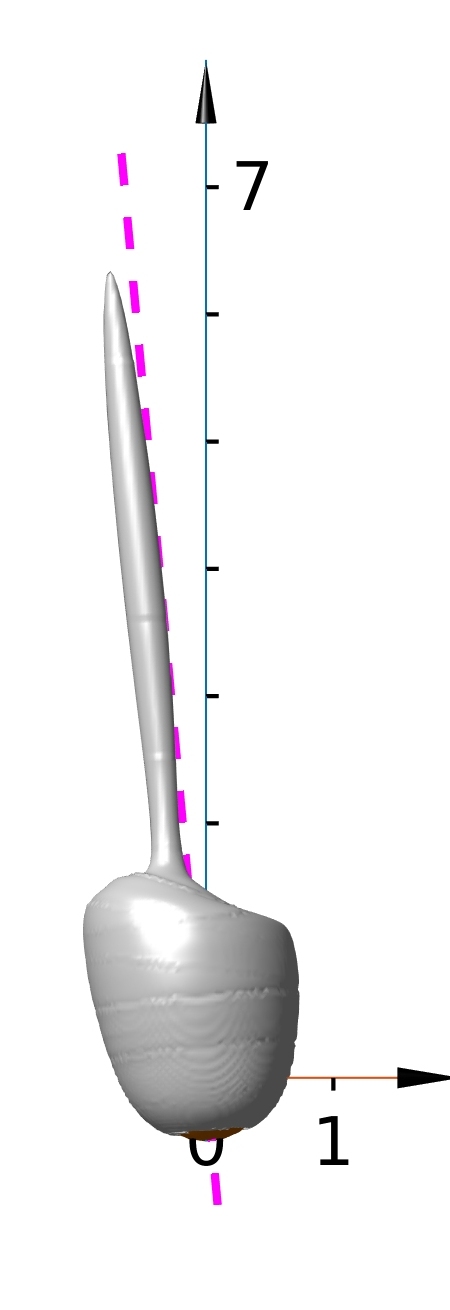}
    \hspace*{-.95\linewidth}\raisebox{2.7\linewidth}{%
      \small$\frac{z_p}{d_p}$
    }
    \hspace*{+.05\linewidth}\raisebox{.1\linewidth}{%
      \small${x_{pH}}/{d_p}$
    }
  \end{minipage}
  \begin{minipage}{.13\linewidth}
    \includegraphics[width=\linewidth]
    {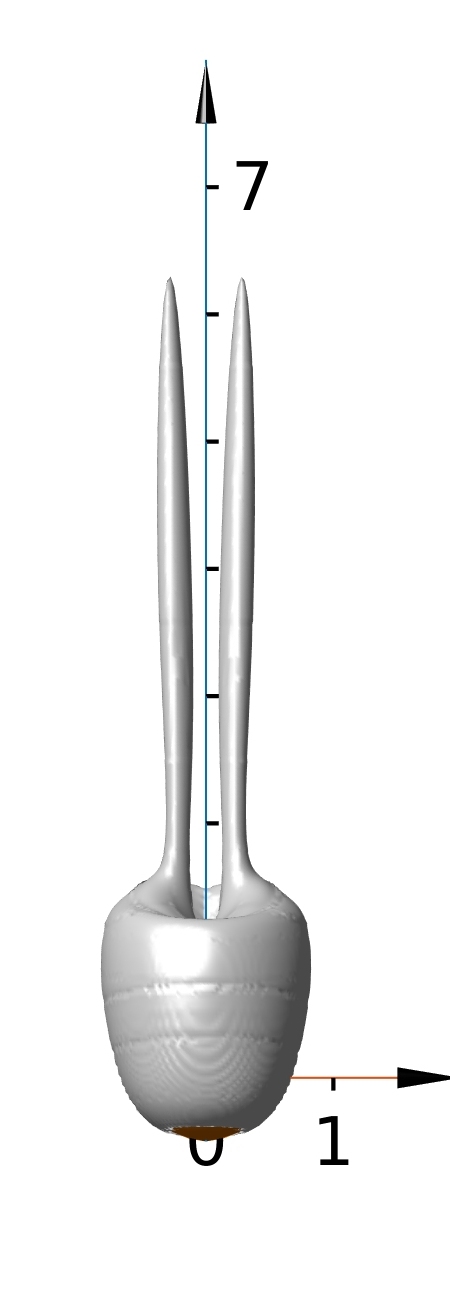}
    \hspace*{-.95\linewidth}\raisebox{2.7\linewidth}{%
      \small$\frac{z_p}{d_p}$
    }
    \hspace*{-.05\linewidth}\raisebox{.1\linewidth}{%
      \small${x_{pHz\perp}}/{d_p}$
    }
  \end{minipage}
  \hfill
  \begin{minipage}{2ex}
    \small$\displaystyle\frac{u_{r\parallel}}{u_g}$
  \end{minipage}
  \begin{minipage}{.3\linewidth}
    \includegraphics[width=\linewidth]%
    {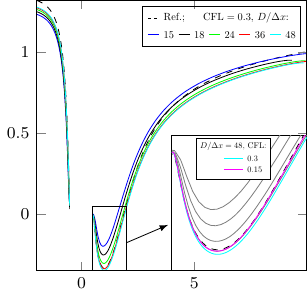}
    \\
    \centerline{\small$x_{p\parallel}/d_p$}
  \end{minipage}
  \hfill
  \begin{minipage}{1ex}
    \small$\displaystyle\frac{u_{pH}}{u_g}$
  \end{minipage}
  \begin{minipage}{.35\linewidth}
    \includegraphics[width=\linewidth]%
    {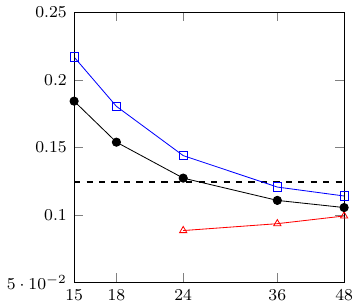}
    \\
    \centerline{\small$d_p/\Delta x$}
  \end{minipage}
  \caption{%
    Flow around a settling sphere in the regime of steady oblique
    motion ($Ga=178.46$, $\rho_p/\rho_f=1.5$) with gravity acting in
    the negative $z$ direction.  
    $(a)$~Lateral and $(b)$ frontal view of the spectral-element
    reference solution \citep{uhlmann:13a} (their case ``BL''),
    showing the double-threaded vortex structure (second invariant of
    the velocity gradient tensor, $\lambda_2=-0.015u_g^2/d_p^2$,
    \citep{jeong:95}).
    $(c)$~Streamwise velocity component of the relative velocity,
    $u_{r\parallel}$, along a line parallel to the trajectory of the 
    particle passing through its centroid (cf.\ the dashed magenta
    line in panel $a$). Solid lines
    correspond to IBM computations with different 
    particle resolutions $d_p/\Delta x$ indicated by colors; 
    the dashed line shows the reference spectral-element
    solution; the inset shows a close-up of the recirculation region,
    and includes a reduction of the time step. 
    $(d)$~The horizontal particle velocity magnitude $u_{pH}$ as
    computed with different resolutions (lines with
    symbols), compared to the reference data from 
    the spectral-element simulation (horizontal dashed line).  
    {\color{black}\solid$\bullet$\solid}~IBM method of
    \citep{uhlmann:04}, $CFL=0.3$; 
    {\color{blue}\solid$\square$\solid}~IBM method of
    \citep{uhlmann:04}, $CFL=0.15$;
    {\color{red}\solid$\triangle$\solid}~LBM computation of
    \citep{rettinger:17}, using central linear interpolation. 
    Note that the IBM data is not identical to that presented in
    \citep{uhlmann:13a}; it has been recomputed here with the
    procedure described in \citep{moriche:20a} such that the
    Galileo number matches the nominal value $Ga=178.46$ exactly.  
  }
  \label{fig:PRM_reference_visu} 
\end{figure} 
The motion of a single particle settling in unbounded ambient flow
already features a rich set of dynamics \citep{ern:12}, where some
regimes of motion exist only in a relatively narrow range of the
parameter space, and therefore single-particle settling constitutes a
stringent test of the numerics.  
Here we focus on a single parameter point which has been proposed as
one of the benchmarks in \citep{uhlmann:13a}: an isolated sphere of 
density ratio $\rho_p/\rho_f=1.5$ is settling at a Galileo number
$Ga=178.46$, which corresponds to the steady oblique regime of motion,
induced by a double-threaded, off-center wake (cf.\ visualization in
figure~\ref{fig:PRM_reference_visu}).
The Galileo number $Ga=u_gd_p/\nu_f$ can be
thought of as a particle Reynolds number computed with the a priori
known gravitational velocity scale
$u_g=((\rho_p/\rho_f-1)d_p|\mathbf{g}|)^{1/2}$ (where $\mathbf{g}$ is
the gravitational acceleration). 
Note that, after possible initial transients, the particle motion and
the flow field are steady in a frame of reference attached to the
particle in this regime, and this is the frame chosen in the reference
computations performed with the method proposed in
\citep{ghidersa:00,jenny:04b}. 
In the lab-frame, however, which is chosen when performing validation
tests with a PR-DNS method, the solution is unsteady. Therefore, the 
capability of the algorithms to capture the relative motion of an
immersed body with respect to the fixed grid can be evaluated. 

Figure~\ref{fig:PRM_reference_visu} shows that the oblique regime is
captured with non-conforming methods on a fixed grid when ${\cal
  O}(10)$ grid points per particle diameter are used. It can also be
seen that the amplitude of the recirculation region is reasonably
well reproduced with $d_p/\Delta x=24$ when using a standard immersed
boundary method \citep{uhlmann:13a}.
The improvement of the prediction of the flow field around the
particle due to refinement yields a corresponding improvement of the
predicted hydrodynamic force, and hence of the computed particle
velocity, as can be seen in figure~\ref{fig:PRM_reference_visu}$(d)$.
In the figure we have included the Lattice-Boltzmann-based results of
\citet{rettinger:17}, which are of similar quality as the IBM results
at comparable spatial and temporal resolutions. Unfortunately, no
direct comparison of the two approaches with respect to computational
cost is available. 
\revision{%
  Note that the two non-conforming methods appear to converge to a slightly
  different value of the lateral particle velocity than the reference
  data, which can probably be attributed to the different lateral
  boundary conditions (periodic vs.\ zero-stress) and to the shape of the
  computational domain (cuboid vs.\ cylindrical). }{%
  Note that the two non-conforming methods appear to converge to slightly
  different values of the particle-related quantities than the reference
  data (e.g.\ the horizontal velocity shown in
  figure~\ref{fig:PRM_reference_visu}$d$, and similarly for the
  vertical component and the angular velocity, which are not shown),
  which can be attributed to the different lateral
  boundary conditions (periodic vs.\ zero-stress) and to the shape of the
  computational domain (cuboid vs.\ cylindrical). 
}
\revision{Note}{Let us mention}
that analogous reference data for a settling oblate spheroid is
available \citep{moriche:20a}, where the authors also discuss the
convergence properties of an immersed boundary method.  
\%
\%
\%
\%
\%
\%
\%
\%
%
%

%% file: chap1_comparison.tex
\revision{}{
\section{%
  \revision{}{%
    Comparing PR-DNS methods: a difficult exercise}
}
\label{sec:PRM-numerics-comp}
There exists a broad spectrum of PRS methods in the literature and
readers may wonder which method to select for their own research. In
the preceeding sections, we elaborated on IBM, DLM and Boltzmann-based
approaches, but we did not discuss many other methods such as ghost
node \cite{mittal:08}, cut-cell \cite{schneiders:16} and overset grid
\cite{henshaw:08} methods for the sake of conciseness. It is 
indeed 
tempting to compare the performance of these different methods and
provide guidelines to readers in their selection process. However,
such a comparison is not straightforward and may lead to erroneous
conclusion if not conducted with care as it involves a variety of
factors. In particular, computing performance is risky to compare as
it deeply depends on programming. In fact, the
same method
implemented by two researchers in two different codes may lead to a
significant performance variation simply because
of implementation details. 
Another source of uncertainty is the
test cases selected to conduct a comparison between assorted
methods. For instance, a sub-set of methods may perform very well in
dilute regimes but not well in dense regimes while another sub-set of
methods may exhibit the reverse performance. Another issue that arises
is the nature of the basic fluid flow solver. While some methods may
be able to share the same fluid flow solver (i.e. the solver in the
absence of suspended particles) such as, e.g., direct forcing/IB and
DLM/FD, other methods are inextricably bound to the basic fluid flow
solver such as, e.g., LBM. While a comparison of PR-DNS methods will
inevitably remain hazardous if not carried out extremely thoroughly,
any research group that would want to attempt
such a comparison should
rely on the following metrics: 
\begin{itemize}
    \item The order of spatial and temporal accuracy of the method.
    \item The magnitude of the spatial error as a function of the grid
      size. Indeed, a method may be of low order, 
      but the prefactor of the error may be very small, thus still rendering
      the method effective. The same applies to the temporal error.   
    \item The scalability on an increasing number of processes
    \item The absolute computing speed measured in terms of number of
      cells resolved per process per time step per second. 
\end{itemize}
PR-DNS methods may also be compared at a more conceptual level through their design properties:
\begin{itemize}
    \item Is the method conservative or non-conservative ?
    \item Is the method easily extendable to adaptive mesh refinement ?
    \item Does the method require hydrodynamic radius calibration and
      consequently does the method easily apply to non-spherical
      shapes ? 
\end{itemize}
But then these properties should again be put to the test to assess
whether some presumably more satisfactory properties (e.g.\ a
conservative method is admittedly more suitable than a
non-conservative method) lead to quantifiable differences in the data
sets and, therefore, to substantially different conclusions on the flow
dynamics when these data sets are analyzed. This is a daunting task
that to the best of the authors' knowledge has not been addressed in
detail by
any previous study.
}

%% file: chap1_conclusions.tex
\section{Conclusion and outlook} 
\label{sec:PRM-conclusion}
In the present chapter we have outlined several approaches for
performing particle-resolved DNS in an efficient manner. This can be
achieved with a non-conforming grid which does not need to be adapted
to the evolving fluid-particle interface. Instead, the appropriate
interface condition (i.e.\ no-slip) is typically imposed by means of
source terms (IBM, cf.\ \S~\ref{sec:PRM-numerics-ibm}), or through
Lagrange multipliers (DLM, cf.\ \S~\ref{sec:PRM-numerics-dlm}) --- 
or possibly with the aid of one of a number of alternative techniques
which have been omitted here for conciseness (e.g.\ sharp interface
method, penalization method, overset grids).
It was also shown that the choice of the basic fluid solver is not
crucial to the success of PR-DNS, and that a wide variety
of discretizations have been successfully employed in the past. 

Numerical methods for PR-DNS have now reached a certain level of maturity,
which is manifested by the relatively large body of literature
reporting results achieved with the aid of this technique (cf.\ 
chapter~6 in the current volume).
It should also be noted that particle shapes other than spheres do not
pose a fundamental challenge to the numerical method, as discussed in 
chapter~11. 
At the same time as the number of degrees of freedom which can be
tackled with this approach are continuously growing, two trends can be
observed. 

First, an additional gain in efficiency might be achieved through the
use of adaptive mesh refinement (AMR), as pioneered in the 
context of IBM by \citet{roma:99}, and extended by other authors
\citep{vanella:14},
as well as in the context of DLM/FD  \citep{selccuk2020fictitious} and
LBM \citep{rohde2006generic,eitel2013lattice,bauer2020walberla}.  
Here the promise is that the benefit in flows with a wide range of
scales separated in space (e.g.\ dilute systems with high particle
Reynolds numbers) will outweigh the higher operation count inherent in
the associated complex data-structures and solvers.
In the future it can be expected that the AMR approach will become more
commonplace as the parameter space for PR-DNS will expand. 

Second, the non-conforming methods discussed here have already 
proven very versatile, and many additional physical processes can 
be added to the purely hydrodynamic problem. A brief and
non-exhaustive list of added physics which have been incorporated into
numerical methods similar to the ones presented in this chapter is
given in the following.  
Heat transfer problems in the fluid domain with different boundary
conditions at the fluid-solid interface can be treated along the same
lines as forcing the momentum equation
\citep{pacheco:05,Yu2006heat,tavassoli:13}. Other authors have developed
approaches for treating conjugate heat transfer problems throughout
the fluid and solid domains \citep{sato:16}.  
Additional forces acting on the particles, beyond those which arise due to
direct solid-solid contact or due to a failure to resolve lubrication
(cf.\ chapter~11), can be integrated into the PR-DNS
framework in a relatively straightforward manner, e.g.\ short-range
cohesive forces \citep{vowinckel:19} and electrical charges
\citep{kang:13}.  
Particle deformability can be realized in the framework of deformable
shell models (applicable to the simulation of capsules) with an
immersed-boundary technique as shown e.g.\ in \citep{krueger:11}.  
Non-conforming methods for PR-DNS can equally be applied to
compressible flow problems involving shock-particle interaction
\citep{brehm:15,vanna:20}.  
As a final point let us comment on the potential for treating phase
changes with the aid of the numerical approaches discussed in the
present chapter. Here the challenge lies in resolving the subtle and
tightly coupled mechanical and thermal balances at the freely-evolving
interface. 
Several authors have shown that liquid-solid melting and
solidification processes can be described based on extensions of the 
general ideas of the methodologies presently discussed, by tracking
Lagrangian marker points which are moving at the local Stefan flow
velocity \citep{huang:14,huang:21}.
Other authors have resorted to a description of the interface
properties based on a (Cahn-Hilliard) phase-field model, coupled to 
immersed-boundary-type discretizations \citep{meng:20}. 
This short glimpse into added physical effects suggests that the
potential for tackling more realistic problems with the aid of the
PR-DNS approach in the future is enormous.
It appears therefore safe to say that we will see a growing number of
applications of this technique to a broader range of systems relevant
to problems found in technical devices and natural environments.
%
%
  %
  %
  %
  %
  %
  %
  %
  %
  %
  %
  %
  %
  %
  %
  %
  %
%